\documentclass[aps,prb,10pt,showpacs,twocolumn,floatfix,superscriptaddress,nofootinbib]{revtex4-1}

\usepackage[nointlimits]{amsmath}
\usepackage{amssymb}
\usepackage{exscale}
\usepackage[T1]{fontenc}
\usepackage{graphicx}
\usepackage{xfrac}

\renewcommand{\appendix}{%
  \clearpage%
  \newpage%
  \par%
  \renewcommand{\thesection}{\Alph{section}}%
  \setcounter{section}{0}%
  \renewcommand{\theequation}{S\arabic{equation}}%
  \setcounter{equation}{0}%
  \renewcommand{\thetable}{S-\Roman{table}}%
  \setcounter{table}{0}%
  \renewcommand{\thefigure}{S-\arabic{figure}}%
  \setcounter{figure}{0}%
  \renewcommand{\thepage}{S-\arabic{page}}%
  \setcounter{page}{1}%
}

\newcommand{\dd}{\mathrm{d}}
\newcommand{\ee}{\mathrm{e}}
\newcommand{\ii}{\mathrm{i}}

\newcommand{\Chi}{\mathrm{X}}

\DeclareMathOperator{\Ai}{Ai}
\DeclareMathOperator{\Tr}{Tr}
\DeclareMathOperator{\arccot}{arccot}
\DeclareMathOperator{\const}{const}
\DeclareMathOperator{\rk}{rk}
\DeclareMathOperator{\sgn}{sgn}

\newcommand{\stack}[2]{\genfrac{}{}{0pt}{}{#1}{#2}}

\begin{document}

\title{Computational Bottlenecks of Quantum Annealing}
\author{Sergey Knysh}
\email{Sergey.I.Knysh@nasa.gov}
\affiliation{QuAIL, NASA Ames Research Center, Moffett Field, California 94035, USA}
\affiliation{SGT Inc., 7701 Greenbelt Rd, Suite 400, Greenbelt, Maryland 20770, USA}

\maketitle

\textbf{%
A promising approach to solving hard binary optimisation problems
is quantum adiabatic annealing (QA) in a transverse magnetic field. 
An instantaneous ground state --- initially a symmetric superposition of
all possible assignments of $N$ qubits --- is closely tracked as it becomes 
more and more localised near the global minimum of the classical energy.
Regions where the energy gap to excited states is small (e.g. at the
phase transition) are the algorithm's bottlenecks.
Here I show how for large problems the complexity becomes 
dominated by $O(\log N)$ bottlenecks inside the spin glass phase, where
the gap scales as a stretched exponential. For smaller $N$, only the
gap at the critical point is relevant, where it scales polynomially, as long
as the phase transition is second order. This phenomenon is demonstrated
rigorously for the two-pattern Gaussian Hopfield Model. Qualitative
comparison with the Sherrington-Kirkpatrick Model leads to similar conclusions.
}
\smallskip

Quantum algorithms offer hope for tackling computer
science problems that are intractable for classical computers \cite{qcbook}. 
However, exponential speed-ups seen in, e.g. number factoring \cite{shor}, 
have not materialised for more difficult NP-complete problems \cite{np}.
Those problems are targeted by the quantum adiabatic annealing algorithm (QA)
\cite{nishimori,farhiX,rmp}. Any NP-hard problem can be recast as
quadratic binary optimisation. QA solves it by implementing a quantum
Hamiltonian, written with the aid of Pauli matrices as
\begin{equation}
  \hat H = - \frac{1}{2} \sum_{i,k=1}^N J_{ik} \\\hat \sigma_i^z \hat \sigma_k^z
  - \sum_{i=1}^N h_i \hat \sigma_i^z - \Gamma(t) \sum_{i=1}^N \hat \sigma_i^x.
  \label{Hann}
\end{equation}
Here the first two terms, diagonal in $z$-basis, encode the objective
function. The last term represents the magnetic field in the
transverse direction, which is decreased from $\Gamma(0) \gg 1$ to 
$\Gamma(T_\text{ann})=0$.
The time $T_\text{ann}$ needed by the algorithm is determined by a
condition that the annealing rate is sufficiently low to inhibit
non-adiabatic transitions:
\begin{equation}
  \dd \Gamma / \dd t \ll \Delta E \cdot \Delta \Gamma.
  \label{dGdt}
\end{equation}
These are most likely near points where the instantaneous gap to
excited states $\Delta E$ attains a minimum as a function of $\Gamma$;
further, $\Delta\Gamma$ is defined as the width of the region where
the gap remains comparable to its minimum value.

QA offers no worst-case guarantees on time complexity \cite{vandam}, 
but initial assessments of \emph{typical case} complexity were optimistic.
Both experimental \cite{brooke} and theoretical \cite{santoro} evidence
hinted at performance improvement over simulated annealing for finite-dimensional
glasses; however, some empirical evidence in support of the theory has recently
been called into question \cite{heim}.
Early exact diagonalisation studies also observed polynomially small
gaps in the constraint satisfaction problem on random hypergraphs
\cite{farhiSci}, but that finding had been challenged by quantum Monte
Carlo studies involving larger sizes \cite{young}.
Benchmarking of a hardware implementation of QA, courtesy of D-Wave
Systems, shows no improvement in the scaling of the performance
\cite{boixo,ronnow}. Whether that might be attributable to a finite
temperature at which the device operates or its intrinsic noise is
unclear at present \cite{boixo2,katzgraber,zhu}.

Statistical physics offer some intuitive guidance: Small gaps develop
at the quantum phase transition point and become exponentially small when
the transition is $1^\text{st}$-order \cite{partitioning,qrem1,qrem2,kxor} or
when different parts of the system become critical at different times
for strong-disorder continuous phase transitions \cite{fisher}.
The most promising candidates for QA are thus problems with bona fide
$2^\text{nd}$-order phase transitions, where the disorder is irrelevant at the QCP.

The scaling analysis described here suggests a polynomially small gap at the
critical point of the archetypal spin glass: the
Sherrington-Kirkpatrick (SK) model \cite{sk1,sk2,sk3}.
It has been pointed out \cite{santoro,altshuler} that QA may still be doomed by
the bottlenecks in the spin glass phase. 
Exponentially small gaps away from the critical point have been
observed in simulations \cite{farhigap},
but adequate theoretical description of this phenomenon has proven challenging.
A perturbative argument in support of this qualitative
picture has been considered in ref.~\onlinecite{altshuler}.
However, the results were not borne out by more accurate analysis
that took into account the extreme value statistics of energy levels
\cite{comment}.

The present manuscript sheds new light on the mechanism of tunneling
bottlenecks in the spin glass phase. Using exact, non-perturbative, methods,
this is illustrated for a simple model, but the main findings are
expected to be valid for quantum annealing of more realistic spin glasses.
During annealing, the system must undergo a cascade of tunnelings
at some specific values of $\Gamma_1,\Gamma_2,\ldots$ in an
approximate geometric progression. 
For a finite system size, these bottlenecks are few, $O(\log N)$,
and may not even appear until $N$ is sufficiently large, highlighting the
challenge of interpreting the results of numerical studies.
Bottlenecks also become increasingly easier as $\Gamma \to 0$,
counter to expectations that tunnelings are inhibited as the model becomes
more classical. A related finding is that the time complexity of QA is
exponential only in some fractional power of problem size: a mild
improvement over more pessimistic estimates \cite{altshuler}.

\section*{Results}
\textbf{Summary.}
The spin glass phase, which is entered below some critical value of the transverse
field $\Gamma_c$, is characterised by a large number of valleys.
Often, this transition is abrupt, driven by competition between an
extended state and a valley (localised state) with the lowest energy, 
as occurs in the random energy model \cite{qrem1,qrem2}.
The exponentially small overlap between the two states then determines
the gap at the phase transition.
However, even if new valleys develop in a continuous manner as
$\Gamma$ decreases, small changes in the transverse field may result
in a chaotic reordering of associated energy levels, leading to
Landau-Zener avoided crossings and attendant exponentially small gaps.

Nonetheless, attempts to make this intuition exact are fraught with
potential pitfalls. For increasing $\Gamma$, two randomly chosen
valleys are equally likely to come either closer together or further
apart in energy.
In the case of the former --- and further, if the sensitivity of energy
levels to changes in the transverse field is so large that the levels
`collide' before either valley disappears ---  avoided crossing will occur.
This may not be necessarily the case when one considers `collisions'
with the ground state, which are of particular concern to QA.
The ground state corresponds to a valley with the lowest energy;
this and other low-lying valleys obey fundamentally different
statistics of the extremes.

A case in point is the analysis of ref.~\onlinecite{altshuler},
which develops perturbation theory in $\Gamma$. The classical limit
($\Gamma=0$) is used as a starting point; how that analysis might be
extended to $\Gamma>0$ has also been discussed \cite{laumann}.
A type of constraint satisfaction problem (CSP) has been considered:
classical energy levels are discrete non-negative integers 
(number of violated constraints) but have exponential degeneracy.
`Zeeman splitting' for $\Gamma>0$ scales as $\sqrt{N}$, which,
for large problems, may be sufficient to overcome the $O(1)$ classical gap and
cause avoided crossings of levels associated with different classical
energies.
Yet this trend disappears if only levels with the smallest energies
(after splitting) are considered; these are relevant for avoided
crossings with the ground state.  
This about-face is not immediately apparent, only coming into play for
$N \gtrsim 100$, when the exponential degeneracy of the classical
ground state sets in. It has, however, been confirmed with analytical
argument and numerics \cite{comment}.
Consequently, arguments based on perturbation theory cannot be used to
establish the phenomenon.

\begin{figure}[th]
  \includegraphics[width=\linewidth]{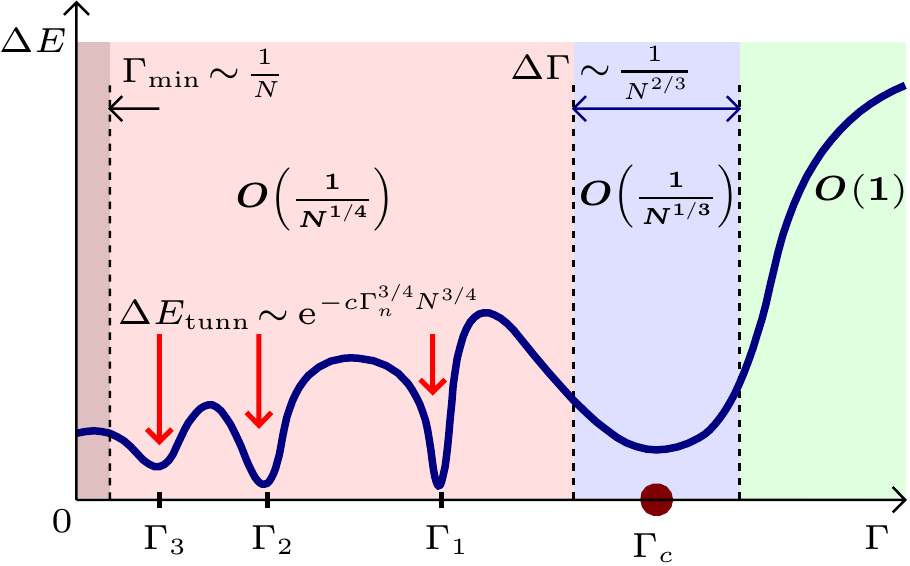}
  \caption{\label{cartoon}
    Sketch of the behavior of the gap in a Hopfield model with the
    Gaussian distribution of disorder variables: 
    in the paramagnetic phase ($\Gamma>\Gamma_c$), in the glassy phase
    ($\Gamma<\Gamma_c$) and in the critical region
    ($\Gamma \approx \Gamma_c$).
    Scaling of the $\emph{typical}$ gap in these regions is indicated
    in bold letters using big-O notation.
    The area
    $\Gamma<\Gamma_\text{min} \sim 1/N$ is
    where the discrete nature of the energy landscape becomes manifest:
    the ground state becomes nearly completely localised.
    The glassy phase also contains $\log N$ isolated bottlenecks,
    indicated with red arrows, where
    the gaps scales as a stretched exponential.}
\end{figure}

This manuscript offers a fresh perspective, illustrated by studying
quantum annealing of the Hopfield model. Mean-field analysis correctly
describes thermodynamics if the number of random `patterns' is small.  
The method is further extended to extract information about exact quantum
energy levels. Importantly, the classical energy landscape is made much
more complex by insisting that the distribution of disorder is Gaussian.
Fig.~\ref{cartoon} sketches a `phase diagram' obtained for this model.
For decreasing $\Gamma$ the gap changes as follows: 
(1) it is finite (does not scale with $N$) in the paramagnetic phase,
$\Gamma>\Gamma_c$;
(2) scales as $1/N^{1/3}$ in the narrow region of width $1/N^{2/3}$
around $\Gamma=\Gamma_c$; 
(3) increases slightly, with typical values scaling as $1/N^{1/4}$ for
$\Gamma < \Gamma_c$.
In addition, there are avoided crossings at isolated points
$\Gamma_1, \Gamma_2, \ldots$, which is a manifestation of
the transverse field chaos, demonstrated conclusively in this work.

The first non-trivial example requires two Gaussian patterns. 
In this case the `energy landscape' is effectively one-dimensional,
which greatly simplifies the analysis. 
The most important element of this analysis is accounting for the
extreme value statistics associated with the valleys (local minima) having
the lowest energies.
To this end, the distribution of energies of the classical landscape must
be conditioned so that they are never below the energy of the global
minimum. This becomes feasible when reformulated as a continuous random
process, in the limit $N \to \infty$.

This particular model is not as interesting from the computer science
perspective, not least because it affords an efficient classical algorithm.
It is sufficiently simple so that a complete quantitative analysis presented 
later on in the manuscript has been possible. Yet, the model captures
the essential properties of the spin glass: its qualitative features directly apply
to much more general models, including Sherrington-Kirkpatrick.
The most important feature of the classical energy landscape is that
exhibits fractal properties, which both ensures that hard bottlenecks
are present in the spin glass phase and also governs their distribution.
The role of the transverse field is to approximately coarse-grain it
on scales determined by $\Gamma$, eliminating small barriers;
thus the number of valleys decreases as $\Gamma$ grows.
A specific random process, corresponding to the energy landscape of
the `infinite'-size instance, will contain every possible realisation of itself 
at some `length scale'. Some realisations will contain high barriers that 
cannot be easily overcome; these will lead to tunneling bottlenecks.

This intuition can be used to immediately establish the scaling of the
number of tunneling bottlenecks. Since the model contains no inherent length
scale in the limit $N \to \infty$, it can be argued that the expected number
of tunneling bottlenecks in a finite interval $[\Gamma_1; \Gamma_2]$ should be
a function of the ratio $\Gamma_2/\Gamma_1$. The logarithm is the only
function that respects additivity, i.e. $\mathcal N_\text{tunn}([\Gamma_1,\Gamma_2]) = 
\mathcal N_\text{tunn}([\Gamma_1;\Gamma'])+\mathcal N_\text{tunn}([\Gamma',\Gamma_2])$.
To obtain the total number of bottlenecks, one considers the interval 
$[\Gamma_\text{min}; \Gamma_c]$. Here $\Gamma_c \sim 1$ is the boundary
of the spin glass phase. The lower cutoff, $\Gamma_\text{min}$, corresponds
to the lowest energy scale of the classical model, which scales as an inverse
power of system size, e.g. as $1/N$ for the Gaussian Hopfield
model. In a sense, tunneling bottlenecks are connected to the $\Gamma=0$
`fixed point' (note that the classical gap vanishes asymptotically).
To summarize, the number of tunneling bottlenecks will grow as
\begin{equation}
  \mathcal N_\text{tunn} \approx \alpha \ln N.
\end{equation}
Locations (in $\Gamma$) will depend on specific disorder realisation,
but self-similarity ensures that the successive ratios $\Gamma_n / \Gamma_{n+1}$ 
converge to a universal distribution.

This logarithmic rise is far weaker than a power law seen in some phenomenological models
of temperature chaos \cite{tchpow} and, as has been argued above, likely
to be a universal feature. The prefactor is model-dependent; its
numerical value can be used to estimate the minimum problem size for
which the mechanism becomes relevant, via $\mathcal N_\text{tunn} \approx 1$.
A value of $\alpha \approx 0.15$ is obtained for the problem at hand,
so that additional bottlenecks become an issue for $N \gtrsim 1000$.
Prior numerical studies similarly required large sizes before the
exponentially small minimum gap was observed \cite{farhigap}, and so far 
there has been no evidence of two or more exponentially small gaps coexisting.
The slow, logarithmic increase of the expected number of bottlenecks
is the most likely culprit.

A notable feature of these results is that tunnelings become
progressively `easier' as $\Gamma \to 0$ despite the fact that the
model becomes more classical. Tunneling gaps increase as
\begin{equation}
  \Delta E_\text{tunn}^{(n)} \sim \ee^{-c \Gamma_n^{3/4} N^{3/4}}.
\label{DEtunn}
\end{equation}
Notice that they cease to be exponentially small for $\Gamma \lesssim
\Gamma_\text{min}$; at that point the ground state is already
localised near the correct global minimum.
The power law exponent for this stretched exponential is model-dependent, 
related to the scaling of barrier heights. These scale as $N^{1/2}$
for the Gaussian Hopfield model, which, together with $O(N)$ scaling of the
effective mass, gives rise to the $N^{3/4}$ term in the exponent.

The finding that the gaps \emph{increase} for smaller $\Gamma$ deserves
explanation. Typically, valleys with similar energies differ by up to
$N/2$ spin flips. This changes once lowest energies are considered:
All spin configurations with energies less than $\epsilon$ above
the global minimum are contained in a neighborhood of radius 
$O\bigl[(N\epsilon)^{2/3}\bigr]$, using Hamming distance as a metric.
The problem is not rendered easy by the mere fact that the global minimum
is so pronounced (although theoretical analysis inspired an efficient classical
algorithm for the $p=2$ Hopfield network, briefly described later on). 
It does imply, however, that the ground state
wavefunction does not jump chaotically: Every subsequent tunneling
involves shorter distances, with $O(\Gamma N)$ spin flips, and achieves
progressively better approximation to the true global minimum. 
Absent such a trend, annealing would be most difficult toward the end
of the algorithm, when $\Gamma \sim 1/N$, and the minimum gap would
exhibit less favorable exponential scaling \cite{altshuler}.

In what follows, the model and its solution are described in greater
detail. First, finite-size scaling of the `easy' QCP bottleneck is linked to
the thermodynamics of the phase transition. The next part goes beyond
thermodynamics, considering small corrections to the extensive
contribution to the free energy. The entire low-energy spectrum,
which depends on disorder realisation, is mapped onto that of
a simple quantum mechanical particle in a random potential.
Finally, extreme value statistics is applied to investigate the
properties of that random potential near its global minimum
by mapping it to a Langevin process. This yields the distribution
of hard bottlenecks in a universal regime ($\Gamma \ll 1$).

\smallskip
\textbf{Quantum Hopfield network.}
Consider a model with rank-$p$ matrix of interactions and no
longitudinal field ($h_i=0$): \cite{hopfield}
\begin{equation}
  J_{ik} = \frac{1}{N} \sum_{\mu=1}^p \xi_i^{(\mu)} \xi_k^{(\mu)}
\end{equation}
(cf. $\rk J_{ik}=N$ for SK model), where $\xi_i^{(\mu)}$ are taken to
be i.i.d. random variables of unit variance.
The thermodynamics of this quantum Hopfield model has been worked out in
great detail by Nishimori and Nonomura \cite{qhm}. 
In fact, that study motivated the development of QA \cite{nishimori}.

When $p$ is small ($J_{ik} \sim 1/N$), it is appropriate to replace local
longitudinal fields with their mean values
$h_i = \sum_k J_{ik} \langle \sigma_k^z \rangle$. The identity
$\langle \sigma_i^z \rangle = h_i / \sqrt{\Gamma^2+h_i^2}$
is used to close this system of equations.
For a transverse field below the critical value, $\Gamma<\Gamma_c=1$,
there appears a non-trivial ($\mathbf{m} \neq 0$) solution to the
self-consistency equation for the macroscopic order parameter, a
vector with $p$ components:
\begin{equation}
  \mathbf{m}= \frac{1}{N} \sum_{i = 1}^N \boldsymbol{\xi}_i \left\langle
  \hat \sigma_i^z \right\rangle.
\label{m}
\end{equation}
Here, the disorder variables are also written using vector notation:
$\boldsymbol\xi_i = \bigl( \xi_i^{(1)},\ldots,\xi_i^{(p)} \bigr)$.

This model is equivalent to the Curie-Weiss (quantum) ferromagnet,
which has a continuous phase transition characterised by a set of
mean-field  critical exponents. Two of these are particularly useful in the
analysis of the annealing complexity: the one for the singular component 
of the ground state energy ($E_0^\text{sing}/N \propto |\gamma|^a$) as 
well as the dynamical exponent for the gap ($E_1-E_0 \propto \gamma^b$),
where $\gamma=\Gamma-\Gamma_c$ is the `distance' from the critical
point.
These exponents are defined for the \emph{infinite} system, yet fairly general
heuristic analysis (see the Methods section) predicts \emph{finite-size} scaling 
for the QCP bottleneck:
\begin{equation}
  \Delta E_c \propto N^{-\frac{b}{a-b}},
  \qquad
  \Delta \Gamma_c \propto N^{-\frac{1}{a-b}}.
  \label{dEdG}
\end{equation}
Substituting values $a=2$ and $b=1/2$ for the problem at hand,
one may estimate the gap at the critical point and the width of the
critical region to be $O(N^{-1/3})$ and $O(N^{-2/3})$ respectively.

Worse-than-any-polynomial complexity of quantum annealing might be
expected for the first order phase transition, which exhibits no
critical scaling (but see ref.~\onlinecite{polyfirst} for an exception).
Another possibility is for the dynamical exponent to diverge at the
infinite randomness QCP: the finite-size gap scales as $\exp\bigl(-c \sqrt{N}\bigr)$
in a random Ising chain \cite{fisher}. For the Hopfield model, however, this scaling
is polynomial, as the disorder is irrelevant at the critical point.
More intriguing is the fact that these pessimistic scenarios are not
found in SK spin glass either: the model is characterised by the same set of
critical exponents, albeit with logarithmic corrections \cite{sk1,sk2,sk3}.
These corrections to scaling increase the gap and, respectively,
decrease the width of the critical region by a factor of $\log^{2/9} N$.
Thus, as long as $T_\text{ann} \gtrsim N$, non-adiabatic transitions
at the critical point should be suppressed. This presents a
conundrum as SK model is known to be an NP-hard problem; finding a
polynomial-time (even in typical case) quantum algorithm would be
a surprising development. The heuristic analysis is clearly insufficient, 
but `digging' deeper into a problem would require a more `microscopic'
analysis. In the following, the problem is mapped to ordinary quantum
mechanics to uncover its low-energy spectrum that explicitly depends
on a particular realisation of disorder, $\{\boldsymbol\xi_i\}$.

\begin{figure}[th]
  \includegraphics[width=0.49\linewidth]{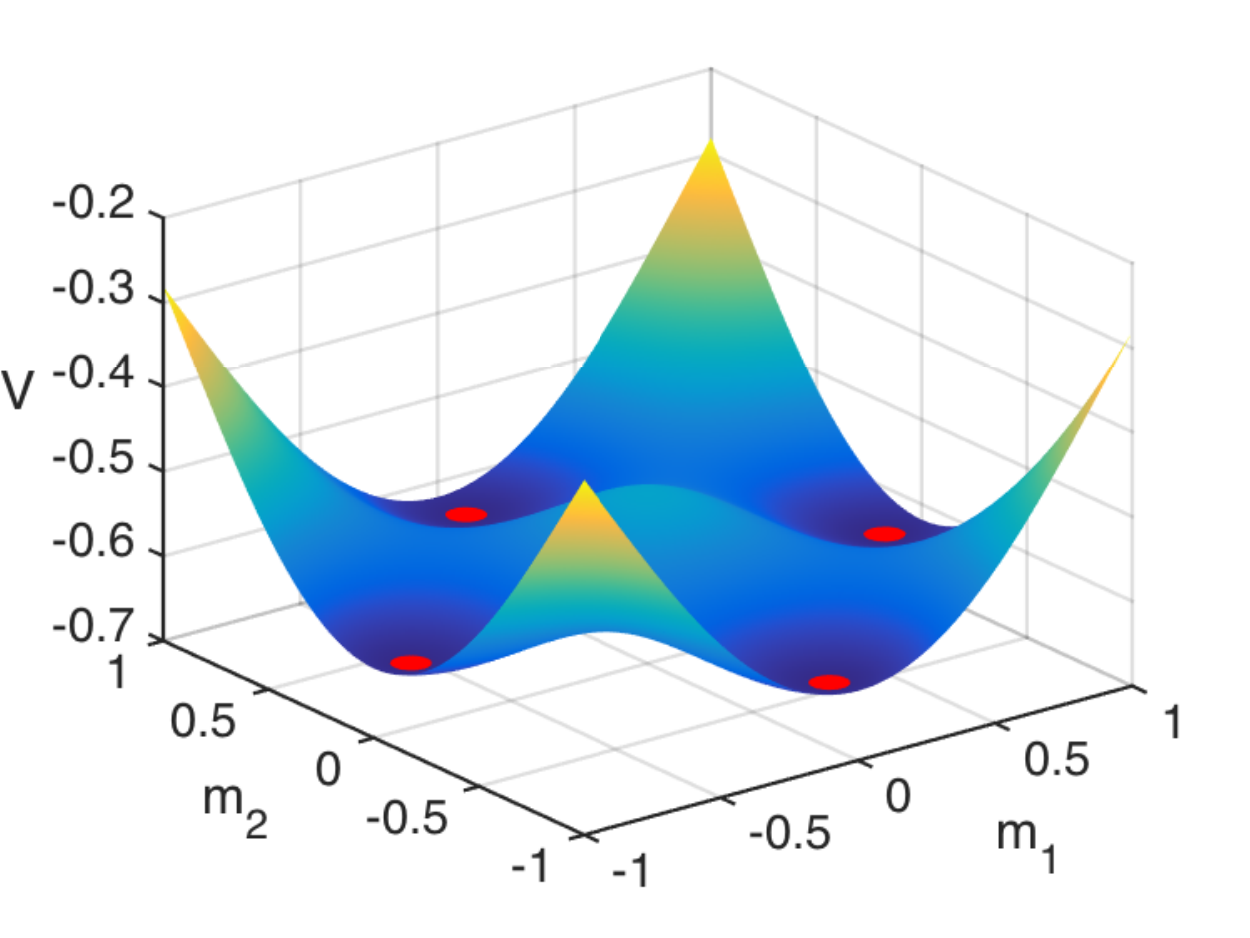}
  \includegraphics[width=0.49\linewidth]{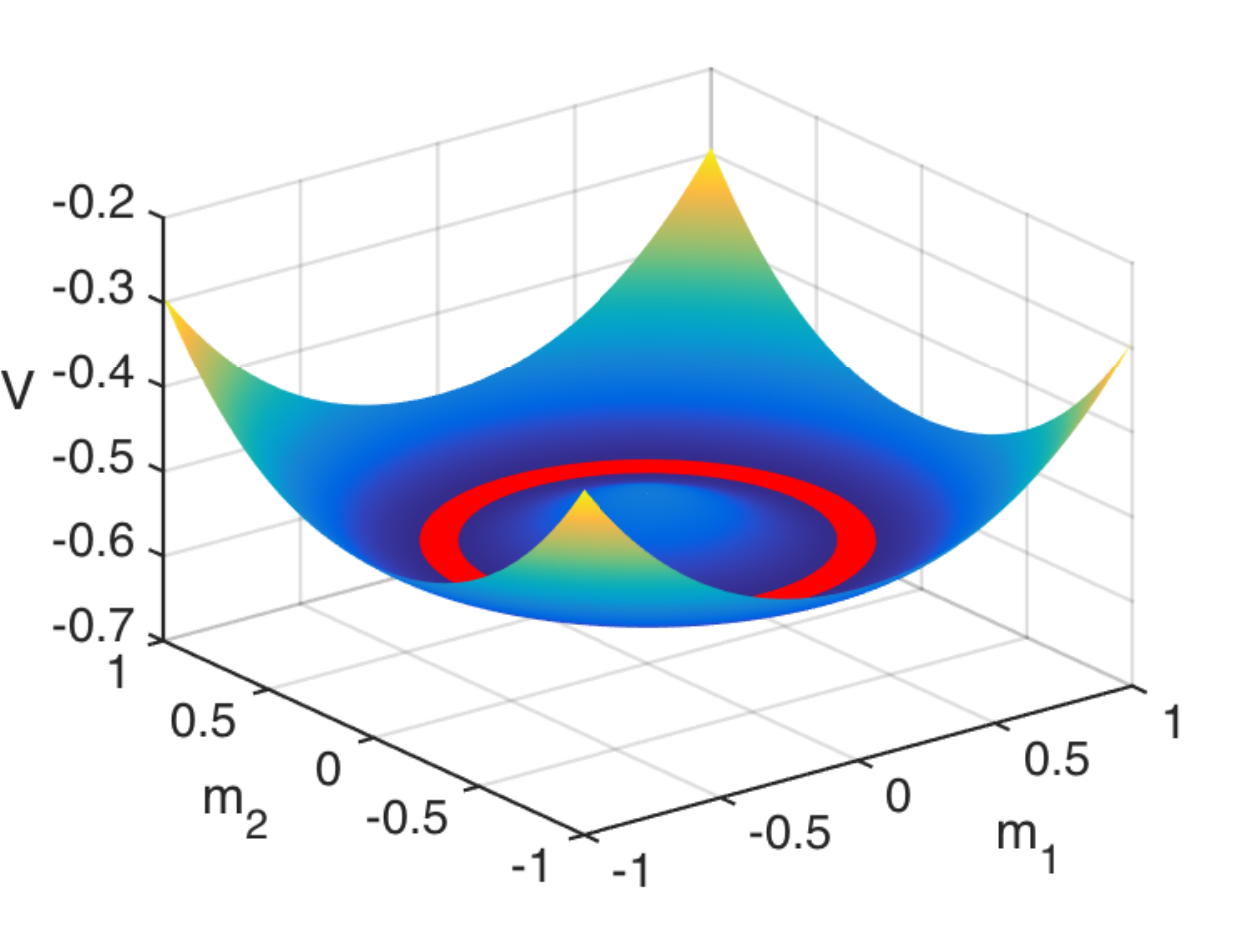}
  \caption{\label{shape}
    The shape of the disorder-averaged effective potential
    $\mathcal{V}(\mathbf{m})$ below the critical point ($\Gamma=0.5<\Gamma_c$)
    for the choice of bimodal (left) and Gaussian (right) distribution
    of disorder variables. Red color highlights the minimum of the potential.
    For the bimodal distribution, the $2p$-fold degenerate global minima 
    are organised in pockets corresponding to encoded patterns. 
    For Gaussian distribution, the degenerate minima
    form a continuum, connected by arbitrary rotations.}
\end{figure}

\smallskip
\textbf{Exact low-energy spectrum.}
Mean-field theory can be derived in a more systematic manner via
Hubbard-Stratonovich transformation. Finite-temperature partition
function $Z(\beta)=\Tr \bigl(\ee^{-\beta H}\bigr)$ can be written as a path
integral over $\mathbf{m}(t)$, which now acquires a dependence on the
imaginary time $0 \leqslant t \leqslant \beta$,  with periodic boundary
conditions: $\mathbf{m}(\beta)=\mathbf{m}(0)$. The value of the
integral is dominated by stationary paths corresponding to the minimum
of an effective potential $\mathcal V(\mathbf{m})$.
While the discussion above has been deliberately equivocal on the
distribution of disorder variables, it is now instructive to contrast 
bimodal ($\xi_i^{(\mu)} = \pm 1$) and Gaussian choices. The shapes of
the effective potential for both scenarios are depicted in Fig.~\ref{shape}.

The conventional bimodal choice defines the model of associative
memory: In the limit $\Gamma=0$ the `patterns' can be perfectly
recalled ($s_i = \pm \xi^{(\mu)}$) when $p$ is small. For the Gaussian
choice, the global minimum corresponds to a mixture \cite{bovier}
\begin{equation}
  s_i = \pm \sgn \biggl( \sum_\mu \alpha_\mu \xi_i^{(\mu)} \biggr),
\end{equation}
rendering memory useless. In the bimodal case, such `spurious' states
only become stable once the number of patterns scales with the problem
size \cite{nnbook}: $p > 0.05N$. The $\textrm{BC}_p$ (bimodal) or $\textrm{O}(p)$
(Gaussian) symmetry of the effective potential is only approximate, to
leading order in $N$. The degeneracy of the ground state is 2
(due to global spin inversion symmetry) for almost all disorder
realisations when $p\geqslant 3$ or $p\geqslant 2$ in the bimodal and
Gaussian scenarios respectively. (Note that that the $p=2$ bimodal case
possesses an additional symmetry, which makes the ground state
4-fold degenerate.)
The system is in a symmetric superposition of the degenerate global
minima at the end of QA. It evolves entirely in the symmetric
subspace since the time-dependent Hamiltonian commutes with 
$\hat U=\exp(\pi \ii \hat S^x)$. Thus, it should be noted that small gaps
between symmetric and antisymmetric wavefunctions are irrelevant to
QA and can be ignored.

Disorder fluctuations `nudge' QA toward the `correct' pattern as it
passes the critical point in the bimodal Hopfield model. No further
bottlenecks are encountered; gaps for $\Gamma<\Gamma_c$ as well as the
`classical' ($\Gamma=0$) gap are $O(1)$. By contrast, the classical
gap scales as $O(1/N)$ in the Gaussian Hopfield model, alerting to a
`danger' posed by the $\Gamma=0$ `fixed point'. To find the low-energy
spectrum when $\Gamma<\Gamma_c$, note that the dominant contribution
to the path integral is from paths where the magnitude of the
`magnetisation' vector remains approximately constant, 
close to its saddle-point value, while the angle is a slow function of time: 
$\mathbf{m}(t) \approx m_\Gamma \bigl( \begin{smallmatrix}
  -\sin \vartheta(t) \\ \cos \vartheta(t) \end{smallmatrix} \bigr)$
for $p=2$. Integrating out the amplitude degrees of freedom, the
partition function is rewritten as
\begin{equation}
  Z(\beta) \propto \int [\dd\vartheta(t)] \ee^{-\int\limits_0^\beta \left(
      \frac{M}{2} \left( \frac{\dd\vartheta}{\dd t} \right)^2 + V_\Gamma(\vartheta)
      \right) \dd t},
\end{equation}
which describes a quantum-mechanical particle of mass $M=O(N)$
moving on a ring, subjected to a random potential
\begin{equation}
  V_\Gamma(\vartheta) = - \sum_i \sqrt{
    \Gamma^2+m_\Gamma^2 \xi_i^2 \sin^2(\vartheta-\theta_i)}
  + N \left< \sqrt{\cdots} \right>_\xi,
  \label{VG}
\end{equation}
where 
$\boldsymbol\xi_i \equiv \xi_i \bigl( \begin{smallmatrix} 
\cos \theta_i \\ \sin \theta_i \end{smallmatrix} \bigr)$ 
and the last term, written in shorthand, adds a constant offset so
that $\langle V(\vartheta) \rangle_{\boldsymbol\xi}=0$. Notice that 
$V_\Gamma(\vartheta) \propto \sqrt{N}$ via central limit theorem,
thereby representing a higher-order correction to the extensive part of the
free energy.

Since the partition function $Z(\beta) = \sum_n \ee^{-\beta E_n}$
encodes information about the spectrum, low-energy (Goldstone)
excitations of the many-body problem are in one-to-one correspondence
with the energy levels of a quantum mechanical particle, up to a
constant shift. The next step is to find the properties of this
potential near a global minimum. These are relevant in a regime
away from the critical point, $\Gamma \ll \Gamma_c$, where the
universal behavior characterised by the appearance of `hard'
bottlenecks sets in.

\smallskip
\textbf{Evolution of the random potential.}
Scaling of the gap for $\Gamma < \Gamma_c$ can be obtained via
semiclassical analysis. Small level splitting due to tunneling between
wells at the two degenerate global minima (this degeneracy is a 
consequence of the global $\mathrm{Z}_2$ symmetry: 
$V_\Gamma(\vartheta+\pi)=V_\Gamma(\vartheta)$) is not relevant to QA.
Higher degeneracies are statistically unlikely; quantisation rules
predict $O\bigl(N^{-1/4}\bigr)$ gaps between energy levels within the
symmetric subspace. But this refers to the \emph{typical} gap,
obtained for \emph{fixed} $\Gamma$ for \emph{almost all} realisations of
disorder. As quantum annealing sweeps the transverse field for a
\emph{fixed} realisation of disorder, $V_\Gamma(\vartheta)$ might
undergo global bifurcation. This would result in a small tunneling gap
for a specific value of $\Gamma$ when the competing minima are in
resonance.

\begin{figure}[th]
  \includegraphics[width=0.49\linewidth]{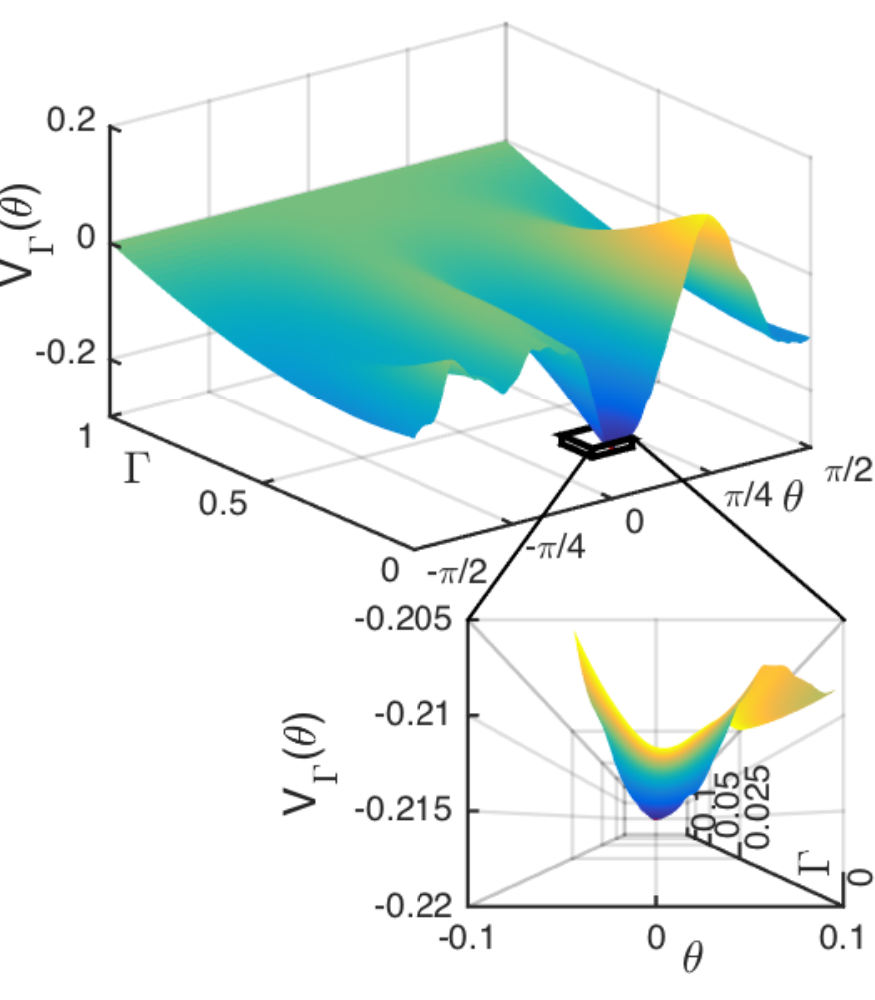}
  \includegraphics[width=0.49\linewidth]{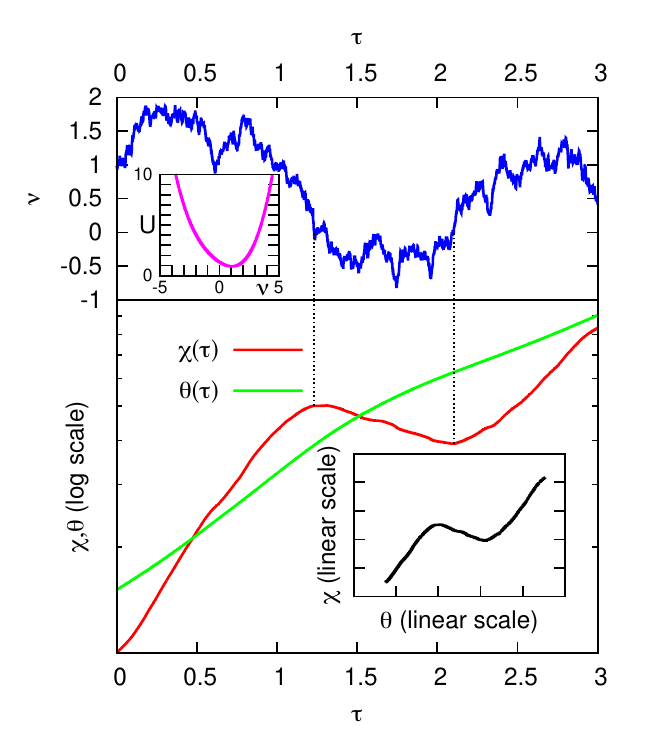}
  \caption{\label{stochastic}
    Illustration of appearance of local minima.
    \emph{Left:} Potential $V_\Gamma(\theta)$ for a specific
    realisation of disorder as a function of $\Gamma$. 
    A perspective-projection 3D plot that zooms on a region near the
    global minimum of $V_0(\theta)$ is shown in the bottom.
    \emph{Right:} Top part of figure plots a specific realisation of 
    a stochastic process $\nu(\tau)$ and a Langevin potential
    $\mathcal U(\nu)$ (inset).
    The bottom part integrates eqn.~\eqref{chith} to obtain
    $\chi(\tau)$ and $\vartheta(\tau)$ in the lower figure. 
    Lower inset plots $\chi$ as a function of $\vartheta$. 
    Regions with $\nu(\tau)<0$ are responsible for the appearance of
    local minima.}
\end{figure}

Such a scenario is impossible near the QCP. Coefficients in the
Fourier expansion of the random potential, 
$\sum_k (a_k \cos 2k\vartheta + b_k \sin 2k\vartheta)$, decrease
as $m^{2k}/\Gamma^{2k-1}$ so that the first harmonic dominates
for $\Gamma \approx \Gamma_c$. Semiclassical analysis confirms a
$O\bigl(N^{-1/3}\bigr)$ gap at the critical point (where the curvature
of the effective potential vanishes, leaving only the quartic part).
As $\Gamma$ decreases, the random potential becomes more rugged
(see Fig.~\ref{stochastic}, left), 
smooth only on scales $\Delta \vartheta \sim \Gamma$, which makes
global bifurcations more likely. Furthermore, it exhibits properties
that allow one to make important predictions without detailed
calculations. Rescaling the potential in the vicinity of either global
minimum $V_\Gamma(\vartheta^\ast)=V_\Gamma^\ast$,
\begin{equation}
  \begin{split}
    \vartheta-\vartheta^\ast & \to \ell (\vartheta-\vartheta^\ast), \\
    V_\Gamma-V_\Gamma^\ast & \to \ell^{3/2} (V_\Gamma-V_\Gamma^\ast),
  \end{split}
  \label{scaling}
\end{equation}
describes the same model but for the rescaled $\Gamma \to \ell\Gamma$ 
and a different realisation of disorder. This scale invariance
is responsible for the logarithmic scaling of the number of tunneling
bottlenecks as has been explained earlier in the text.
However, it still remains necessary to demonstrate that the density of bottlenecks 
is \emph{non-zero}, which entails an examination of the properties of the random
potential in the limit $\Gamma=0$.

\begin{figure}[th]
\includegraphics[width=\linewidth]{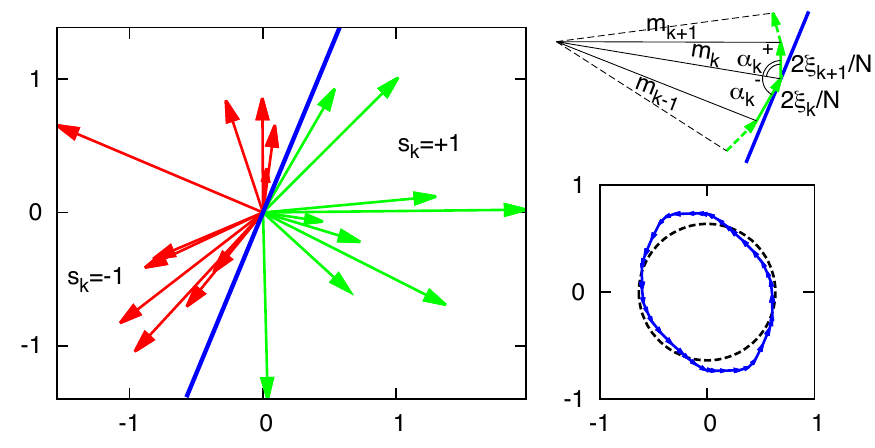}
\caption{\label{geom}
  Geometric interpretation of the random potential, illustrated for a
  random instance with $N=20$. Vector $\mathbf{m}(\vartheta)$ is
  defined by drawing a separating line at angle $\vartheta$ and
  assigning spins with $\boldsymbol\xi_i$ on either side of separating
  line values $s_i=+1$ and $s_i=-1$ respectively (left figure).
  For changing $\vartheta$ vector $\mathbf{m}(\vartheta)$ is
  incremented/decremented by $(2/N)\boldsymbol\xi_i$ (top right).
  The increments form a closed path, approximated by a circle;
  fluctuations define a random potential (bottom right).}
\end{figure}

The classical optimisation problem corresponds to maximising the magnitude
of $\sum_i \boldsymbol\xi_i s_i$. A necessary condition for a local
minimum is that two sets of vectors, $\{\boldsymbol\xi_i|s_i=+1\}$ and
$\{\boldsymbol\xi_i|s_i=-1\}$, can be separated by a line. 
As the angle of this line changes, fluctuations of the amplitude
give rise to a random potential $V_0(\vartheta)$ (see Fig.~\ref{geom}).
This suggests a linear-time algorithm for finding a global minimum:
Sort all vectors by angle (this may introduce an extra $\log N$ factor
due to sorting overhead) and exhaustively check all possible angles.
Of course, QA algorithm is too generic to exploit the specific structure 
of the problem; moreover \emph{ad hoc} efficient algorithms are unlikely
to exist for more general spin glass problems.

On short intervals, the random process is described as an undamped
Langevin process \cite{sinai,groeneboom} in the continuous ($N \to
\infty$) limit (hence the exponent of $3/2$ in eqn.~\eqref{scaling},
corresponding to its fractal dimension). Properly taking into account
the statistics of the extremal properties, the process must be
conditioned on the fact that $V_0(\vartheta) > V_0(\vartheta_0^\ast) =
V_0^\ast$ away from the global minimum. As described in Methods and
the Supplement, such a conditioned process consists of two uncorrelated
branches, $V_0(\vartheta)-V_0^\ast \propto \sqrt{N} \, \Bigl\{
\begin{smallmatrix}
  \chi_+, & \vartheta>\vartheta_0^\ast, \\
  \chi_-, & \vartheta<\vartheta_0^\ast.
\end{smallmatrix}$
Integrating equations
\begin{equation}
  \begin{split}
    \dd(\ln \chi_\pm)/\dd \tau &= \nu_\pm(\tau), \\
    \dd \vartheta/\dd \tau &= \pm\chi_\pm^{2/3},
  \end{split}
  \label{chith}
\end{equation}
defines $\chi_+(\vartheta)$ and $\chi_-(\vartheta)$ parametrically,
in terms of random processes $\nu_\pm(\tau)$ that correspond to a
Brownian motion in a non-linear potential depicted in Fig.~\ref{stochastic}.
This potential is biased toward positive `velocities' $\nu$ so that
$V_0-V_0^\ast \sim \sqrt{N} (\vartheta-\vartheta^\ast)^{3/2}$ from
eqn.~\eqref{chith}. It will, however, experience arbitrary percentage
drops due to subpaths with $\nu<0$ (albeit with decreasing
probability). 

\begin{figure}[th]
  \includegraphics{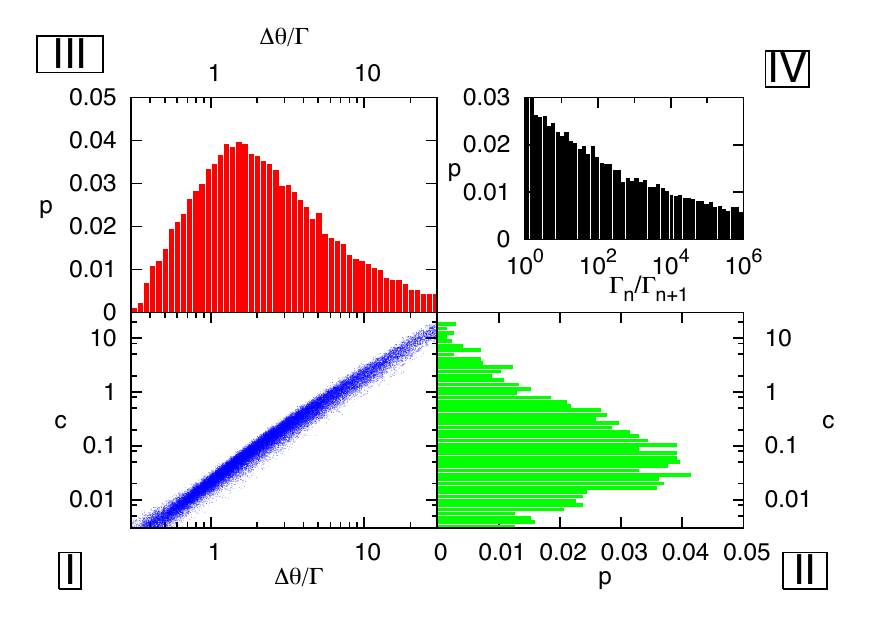}
  \caption{\label{num}
    Numerical results. The L-shaped figure (composed of panels I-III)
    plots the distribution of the tunneling jumps and the exponent. 
    The lower left quadrant (panel I) is a scatterplot
    of $\Delta \vartheta / \Gamma$ (proportional to the number of spin
    flips normalised by $\Gamma N$) against 
    $|\log \Delta E_\text{tunn}|/(\Gamma N)^{3/4}$ (the prefactor $c$)
    for all tunneling events.
    Histograms of these quantities alone (projections of the
    scatterplot) are in the bottom right (panel II) and the top left (panel III) quadrants.
    The top right quadrant (panel IV) plots a histogram of distribution of $\Gamma_n /
    \Gamma_{n+1}$ for the successive tunneling events.}
\end{figure}

For small but finite $\Gamma$, the `separating line'  becomes blurred. 
The random potential adds a `quantum correction' (see Methods and the Supplement),
$V_\Gamma(\vartheta)-V_0(\vartheta)=O\bigl(\sqrt{N}\Gamma^{3/2}\bigr)$. 
For two minima to come into resonance, they cannot be more than
$\Delta\vartheta \sim \Gamma$ apart (i.e. $O(N\Gamma)$ spin flips).
The tunneling exponent is given by under-the-barrier action
$\mathcal A \sim \sqrt{M\Delta V} \Delta \vartheta$, where the
effective mass $M \sim N/\Gamma^2$ and $\Delta V \sim \sqrt{N} \Gamma^{3/2}$,
leading to eqn.~\eqref{DEtunn}.
Numerical results for the universal distribution of tunneling jumps
and the tunneling exponents are exhibited in Fig.~\ref{num}.

\section*{Discussion}
Poor scalability of classical annealing of spin-glass models had been
linked to the phenomenon of temperature chaos \cite{chaos}. Interestingly,
its existence in mean-field glasses had been debated \cite{tchsk1,rizzo,tchsk2},
although it is uncontroversial in finite-dimensional models \cite{tchea1,tchea2}.
Similarly, failures of quantum annealing might be attributed to \emph{transverse field chaos}.
The phenomenon described in this manuscript represents a much stronger finding. 
The mere fact that ground states at $\Gamma$ and $\Gamma+\Delta\Gamma$ will
have vanishingly small overlap as $N \to \infty$ is not inconsistent
with the continuously evolving ground state and poses no `threat' to QA. 
By contrast, `hard' bottlenecks correspond to isolated discontinuities
that persist as $\Delta\Gamma \to 0$.

To dwell upon the generality of these results, first note that scaling
of the tunneling exponent will depend on the universality class of the
model. The SK model, for instance, exhibits different scaling of
barrier heights, believed to be $\propto N^{1/3}$ (see e.g. ref.~\onlinecite{barriers}).
In the model studied, the decrease in the number of spins involved in
tunnelings offsets the divergence of the effective mass in the classical limit
$\Gamma \to 0$. As the height of the barriers also decreases,
the tunneling gaps widen toward the end of the algorithm.
One might expect qualitatively similar behavior in realistic spin glasses.

The logarithmic scaling of the number of bottlenecks is due to
self-similar properties of the free energy landscape in the interval
$[\Gamma_\text{min},\Gamma_c]$. The lower cutoff should correspond to
the smallest energy scale in the classical limit, which for the SK model
is also a negative power of $N$, namely $1/\sqrt{N}$. This is related to the
linear vanishing of the density of distribution of effective fields as
$h \to 0$ at zero temperature \cite{h0} (since $\int_0^{h_\text{min}} P(h) \dd h \sim 1/N$).
The picture is less clear for constraint satisfaction problems, where
the energies are constrained to be non-negative integers; i.e. the
classical gap is $O(1)$. These energy levels have exponential degeneracy, 
which is lifted by the transverse field. 
A value sufficient to make the spectrum effectively quasi-continuous
might  serve as an appropriate lower cutoff $\Gamma_\text{min}$ in problems of this type.
Lack of `hard' bottlenecks in the Hopfield model with the bimodal distribution of disorder
and $p=O(1)$ could be attributed to the fact that the number of valleys is
finite, which is not representative of a true spin glass.

An interesting observation is that since `hard' bottlenecks correspond
to Landau-Zener crossings, annealing times need not be exponentially small.
The probability that QA fails to follow the ground state every single
time in $n$ repeated experiments is
\begin{equation}
  p_\textrm{fail}^{(n)} = \Bigl( \ee^{-\frac{\pi\, \Delta E \cdot \Delta\Gamma}{4 (\dd \Gamma / \dd t)}} \Bigr)^n,
\end{equation}
which exactly matches the probability of failure for the annealing rate
that is $n$ times slower. Using shorter annealing cycles with many repetitions
can minimize the effects of decoherence. It suffices that non-adiabatic 
transitions are suppressed at the critical point only, $\frac{\dd\Gamma}{\dd t} \lesssim 1/N$.

Even with polynomial annealing rates, coherent evolution would require
much better isolation from the environment than what is currently feasible.
The only commercial implementation of QA (D-Wave) must contend with a
fairly strong coupling to a thermal bath. On the positive side, it accommodates 
faster annealing cycles, acting as a `safety valve' to dissipate any heat generated
during the non-adiabatic process. At the same time, it all but ensures that the system 
is always in thermal equilibrium with the environment. 

The theory presented here breaks down when $\Gamma<T$ so that quantum
bottlenecks described here may not be a limiting factor if $\ln(\Gamma_c/T) \lesssim 1/\alpha$.
Main source of errors will be from exponentially many thermally occupied excited states.
If the annealing profile were adjusted so that the energy spacing
increased beyond $T$ towards the end, this would effectively
implement classical annealing. An intricate relationship between temperature, 
problem size, and the properties of the spin glass model might determine 
which mechanism (quantum or classical annealing) will be dominant.

Yet another tradeoff in the design of D-Wave chip is a quasi-2D topology 
of interactions $J_{ik}$ due to fabrication constraints, which incurs significant
performance penalty when mean-field models are `embedded' into a `Chimera' graph
\cite{venturelli}. So-called `native' problems corresponding to uniformly random
instances on this quasi-2D lattice have been argued to be poor candidates for QA
due to the lack of a finite-temperature classical phase transition \cite{chimeraT0};
at the same time, a quantum phase transition at $\Gamma_c>0$ is expected in 2D
glasses \cite{quantum2d}.

Whereas $1^\text{st}$-order phase transition immediately implies
exponential complexity, even for small sizes, problems having a
continuous phase transition may remain tractable up to a threshold, $N_c$,
beyond which tunneling bottlenecks become dominant ($\alpha \ln N_c \sim 1$).
This `tractability threshold' serves as a silver lining fot this otherwise negative result. 
Moreover, the picture of `hard' bottlenecks may be equally applicable
to classical annealing. While in some crafted examples classical annealing
is at a unique disadvantage due to $1^\text{st}$-order phase transition \cite{farhisa},
for most interesting problems both classical and quantum transitions are
$2^\text{nd}$-order. In such a scenario, the density of bottlenecks
becomes a tie-breaker for evaluating relative performance. Whether
quantum annealing can be advantageous in terms of this metric and determining 
which models will benefit is a practically important question for follow-up work.

There remains a question whether the failure mechanism described here can be 
somehow circumvented. Such a feat is feasible, e.g. for a disordered Ising chain,
where an exponentially small gap develops at the critical point, which is
a manifestation of Griffiths singularities \cite{fisher}. Modification of QA that
requires controlling the transverse field for each spin individually can
suppress these singularities and restore a polynomial gap.

Based on comparison with another exactly solvable model, it seems that frustration
--- in addition to disorder --- is essential for the appearance of small gaps 
in the spin glass phase proper. The seemingly random profile of the energy landscape
for finite $\Gamma$ heralds difficulties in avoiding these bottlenecks in generic 
spin glasses. Although any prospects of exponential speedup should be met with
skepticism, heuristics inspired by spin glass theory revolutionalised
branch-and-bound algorithms for constraint satisfaction problems \cite{algo}.
One can remain hopeful that theoretical advances can similarly aid quantum optimisation.

\section*{Methods}

\begin{footnotesize}

Finite-size scaling of the gap at QCP is best understood using an
example of a finite-dimensional system. In thermodynamic limit,
both correlation length and characteristic time diverge near the
phase transition as
\begin{equation}
  \xi_c \propto (\Gamma-\Gamma_c)^{-\nu}, \qquad
  \tau_c \propto (\Gamma-\Gamma_c)^{-z\nu}.
\end{equation}
In a finite system this divergence is smoothed out as soon as the
correlation length becomes comparable to lattice size ($\xi_c \sim N^{1/d}$).
The minimum gap (the reciprocal of $\tau$) and the width of the
critical region should scale as $N^{-z/d}$ and $N^{-1/(d\nu)}$ respectively. 
In this paper, the product $z\nu$ has been labeled as exponent $b$. 
Singular behavior of normalised ground state energy (free energy) is
related to the specific heat exponent ($a=2-\alpha$). Dimensionality
$d$ can be eliminated with the aid of hyperscaling relation
$2-\alpha=(d+z)\nu$ to yield eqn.~\eqref{dEdG} of the main text. 
Independent estimates of the specific heat exponent can be
obtained from the exponents for the order parameter and the susceptibility
($2-\alpha=2\beta+\gamma$).

Finite-temperature partition function can be written as a sum over
a set of paths $[s_i(t)]$ with $0 \leqslant t \leqslant \beta$, 
where $s_i(t)$ alternates between the values $\pm 1$. 
Hubbard-Stratonovich transformation can be used to rewrite it as a
path integral
\begin{equation}
  \begin{split}
    Z &\sim \sum_{[\{s_i(t)\}]} \ee^{\frac{1}{2N} \int_0^{\beta} 
      \left( \sum_i \boldsymbol\xi_i s_i(t) \right)^2 \dd t + 
      \sum_i \mathcal K_\Gamma[s_i(t)]} \\
    & \sim \int[\dd \mathbf{m}(t)] \ee^{-\frac{N}{2} \int_0^{\beta} 
      \mathbf{m}^2(t) \dd t + \sum_i \ln Z_i}.
  \end{split}
  \label{Z}
\end{equation}
The `kinetic' term $\mathcal K_\Gamma[s(t)]$ in the first equation penalises
kinks, representing $\Gamma$-dependent ferromagnetic couplings between
Trotter slices. As the interaction term is decoupled, the problem
reduces to that of evaluating the single-site partition function $Z_i$ for a spin
subjected to a magnetic field with the transverse component $\Gamma$
and the `time-dependent' longitudinal component $h_i(t)=\boldsymbol\xi_i
\mathbf{m}(t)$.
To leading order in $N$, the saddle-point of the path integral \eqref{Z}
is a solution of mean-field equations. This becomes a degenerate
manifold for Gaussian disorder distribution; to determine higher-order
contribution all paths such that $|m(t)|=m_\Gamma$ are considered. 
Evaluating $Z_i$ is best performed in the adiabatic basis that
diagonalises the 2-level Hamiltonian $\hat H_i=-h_i(t)\hat\sigma^z-
\Gamma\hat\sigma^x$,
\begin{equation}
  Z_i = \Tr \mathcal{T} \ee^{\int_0^\beta \hat H_i(t)} \dd t
  = \Tr \mathcal{T} \ee^{-\int_0^\beta\left(\hat E_i(t)
      +\ii\frac{\partial}{\partial t} \hat V_i(t) \right) \dd t}.
  \label{Zi}
\end{equation}
Here $\hat E_i(t)$ is diagonal with eigenvalues
$\pm\sqrt{\Gamma^2+h_i^2(t)}$. Its fluctuations around the mean give
rise to the random potential $V_\Gamma(\vartheta)$.
Non-adiabatic terms due to rotation of the basis ($\hat U=\ee^{-\ii\hat V_i(t)}$)
are treated using 2nd order perturbation theory, giving rise to a kinetic
term $\propto (\dd\vartheta/\dd t)^2$. Note a simple form of the
perturbative term in eqn.~\eqref{Zi} owing to the fact that $\hat V_i(t)$
commute for all $t$.

In the limit $N\to\infty$, the random potential $V_\Gamma(\vartheta)$
defined in eqn.~\eqref{VG} in the main text converges to a Gaussian
process that can be specified completely by its covariance matrix
$\bigl< V_\Gamma(\vartheta) V_{\Gamma'}(\vartheta') \bigr>$.
This can be `diagonalised', alternatively expressing the random
potential as a linear combination of white-noise processes
$\{\zeta_n(\vartheta)\}$.
One representation, as a convolution series $\sum_{n=0}^\infty
\bigl( f_\Gamma^{(n)} \!\star \zeta_n \bigr) (\vartheta)$ with kernels
\begin{equation}
  f_\Gamma^{(n)}(\vartheta) \propto \int_0^\infty
  \sqrt{\Gamma^2+\xi^2 m_\Gamma^2 \sin^2\vartheta}
  \; \xi^{\alpha+1} \ee^{-\xi^2/2} L_n^{(\alpha)}\Bigl(\tfrac{\xi^2}{2}\Bigr) \dd \xi,
\end{equation}
relies on orthogonal properties of associated Laguerre polynomials.
The choice $\alpha=1$ ensures that only $n=0$ term survives in the
limit $\Gamma=0$. The random potential should satisfy a stochastic
equation
\begin{equation}
  \frac{\dd^2 V_0}{\dd \vartheta^2} + V_0 =
  \frac{2\sqrt{N}}{\pi} \zeta_0.
  \label{Vzeta}
\end{equation}
As a side note observe that a similar equation is obtained by taking a
continuous limit in the identities that follow from elementary geometry
(see Fig.~\ref{geom} in the main text, $V_k = \frac{N}{2}m_k^2$):
\begin{multline}
  \tfrac{V_{k+1}-V_k}{\xi_{k+1}} - \tfrac{V_k-V_{k-1}}{\xi_k} = \\
  \tfrac{2}{N}(\xi_{k+1}+\xi_k) + 4\sqrt{2V_k/N}
  \cos \tfrac{\alpha_k^+-\alpha_k^-}{2} \cos \tfrac{\alpha_k^++\alpha_k^-}{2}
\end{multline}

For finite $\Gamma$, the form of the potential is modified as follows:
It is convolved with a smoothing kernel of width $\Delta\vartheta \sim
\Gamma$, which raises the global minimum by $O\bigl( \sqrt{N}
\Gamma^{3/2} \bigr)$. Additional random contributions (from
$n\geqslant 1$) have similar scaling.

In the vicinity of the global minimum, the statistics of the classical
potential is fundamentally different. The `returning force' in
eqn.~\eqref{Vzeta} can be neglected; additionally the process should
be conditioned on the fact that it stays above its value at $\vartheta=0$
(no generality is lost by choosing the global minimum as the origin).
This problem has been a subject of a considerable body of work 
\cite{sinai,groeneboom}, although important aspects ought to be revisited. 
Here, I present a particularly simple self-contained derivation.

A pair $(\chi,\upsilon)$ --- where $\chi \propto V_0-V_0^\ast$ is
interpreted as the `coordinate' and $\upsilon=\dd\chi/\dd\vartheta$
as the `velocity' ($\vartheta$ being `time') --- forms a Markov process.
The probability distribution $p(\vartheta;\chi,\upsilon)$ satisfies,
for $\vartheta>0$,
\begin{equation}
  \frac{\partial p}{\partial \vartheta} = 
  - \upsilon \frac{\partial p}{\partial \chi} 
  + \frac{1}{2} \frac{\partial^2 p}{\partial \upsilon^2},
  \quad \text{while } p(\vartheta;+0,\upsilon)=0\;
  (\forall\upsilon>0)
  \label{pde}
\end{equation}
serves as a boundary condition for the absorbing boundary.
This probability is `renormalised' to condition on the fact
that it survives until some $\Theta \gg \vartheta$. 
It becomes a conserved quantity, but the diffusion equation adds
a drift, $-\partial (\log P_\Theta)/\partial\upsilon$, repelling
from the boundary. The `survival' probability $P_\Theta$ in the limit
$\Theta\to\infty$ is, up to `time-reversal', the universal asymptotic
solution of \eqref{pde}, reduced to ordinary differential equation
(ODE) using scaling ansatz:
\begin{equation}
  p(\vartheta,\chi,\upsilon) \propto 
  \frac{\chi^{2\alpha/3} p_\ast\bigl( \upsilon\big/\chi^{1/3}\bigr)}{\vartheta^\alpha}.
\end{equation}
This exploits a fact that fractal dimensions of `velocity' and
`coordinate' are $[\upsilon]=[\vartheta]^{1/2}$ and
$[\chi]=[\vartheta]^{3/2}$. The asymptotics is dominated by solutions with
the smallest possible exponent, $\alpha=1/4$ out of the infinite set
of eigenvalues for the ODE. This matches a known value obtained with
a different method \cite{sinai}.

The next step performs a change of variables, introducing 
`dimensionless' velocity $\nu=\upsilon/\chi^{1/3}$, and `logarithmic'
coordinate $\mu=\ln \chi$. With the `time' variable redefined via
$\dd\tau=\dd\vartheta/\chi^{2/3}$, Markov process is described by
a tuple $(\vartheta,\mu,\nu)$. Marginalising out $\mu$ and
$\vartheta$ in the equation for $p(\tau;\vartheta,\mu,\nu)$ produces
Fokker-Planck equation for a stochastic motion of particle in a potential
\begin{equation}
  \mathcal{U}(\nu)=- \ln \left(
  \bigl(\nu/6^{1/3}\bigr) \Ai \bigl(\nu^2/6^{2/3}\bigr) + 
  \Ai'\bigl(\nu^2/6^{2/3}\bigr) \right).
\end{equation}
Given a particular realisation of $\nu(\tau)$, the full process
$(\mu,\nu,\vartheta)$ is determined deterministically, by integration
(see eqn.~\eqref{chith} in the main text). The construction of a
realisation of a random process is performed independently for
$\vartheta>0$ and $\vartheta<0$. 

Numerical simulations rescale this random potential instead of
evolving the transverse field: $\vartheta \mapsto
\ee^{-\frac{\Delta\Gamma}{\Gamma}} \vartheta$ 
and $\chi \mapsto \ee^{-\frac{3}{2} \frac{\Delta\Gamma}{\Gamma}} \chi$. 
The process is extended to larger values of $\tau$ as needed (details
of the process for small $\tau$, where they fall below the numerical
precision, are `forgotten'). A fairly large range of $\tau$ is required
to gather the sufficient statistics.

The code used for simulations and the raw numerical results are 
available upon request.

\end{footnotesize}

\begin{footnotesize}

\section*{Acknowledgements}
I would like to thank Vadim Smelyanskiy for useful discussions.
This work was supported in part by the Office of the Director of
National Intelligence (ODNI), Intelligence Advanced Research Projects
Activity (IARPA), via IAA 145483, and by the Air Force Research
Laboratory (AFRL) Information Directorate under grant F4HBKC4162G001. 
The views and conclusions contained herein are those of the author and
should not be interpreted as necessarily representing the official
policies or endorsements, either expressed or implied, of ODNI, IARPA,
AFRL, or the U.S. Government. 
The U.S. Government is authorised to reproduce and distribute reprints
for Governmental purpose notwithstanding any copyright annotation thereon. 

\end{footnotesize}

\newpage
\appendix

\onecolumngrid
\section*{\selectfont\Large Supplementary Information}
\twocolumngrid

\begin{footnotesize}

\textbf{Mathematical Preliminaries.}
This paper makes extensive use of Meijer G-function defined via
intergal
\begin{multline}
  G_{p,q}^{m,n} \biggl( z \bigg| \stack{a_1,\ldots,a_n; a_{n+1},\ldots,a_p}
  {b_1,\ldots,b_m; b_{m+1},\ldots,b_q} \biggr) = \\
  \frac{1}{2\pi\ii} \int_\mathcal{L} \frac{
    \prod_{k=1}^m\boldsymbol\Gamma(b_k+s) \,
    \prod_{k=1}^n\boldsymbol\Gamma(1-a_k-s)}{
    \prod_{k=n+1}^p\boldsymbol\Gamma(a_k+s) \,
    \prod_{k=m+1}^q\boldsymbol\Gamma(1-b_k-s)}
  z^{-s} \dd z,
\end{multline}
where the contour $\mathcal{L}$ is chosen appropriately. 
Bold $\boldsymbol\Gamma(x)$ is used to denote $\Gamma$-function to avoid
confusion with the transverse field variable. G-function can be used
to express many elementary and special functions; those used here
include:
\begin{itemize}
  \item $\ee^{-x} = G_{0,1}^{1,0} \bigl(x \big| \stack{-}{0} \bigr)$,
  \item $(1+x)^{-a} = \frac{1}{\boldsymbol\Gamma(a)}
  G_{1,1}^{1,1} \bigl( x \big| \stack{1-a}{0} \bigr)$,
  \item $L_n^{(\alpha)}(x) \ee^{-x} = \frac{1}{n!} 
    G_{1,2}^{1,1} \bigl( x \big| \stack{-n-\alpha}{0;-\alpha} \bigr)$,
  \item $U(a,b,x) = 
    \frac{1}{\boldsymbol\Gamma(a)\boldsymbol\Gamma(a-b+1)}
    G_{1,2}^{2,1} \bigl( x \bigl| \stack{1-a}{0,1-b} \bigr)$,
\end{itemize}
where $L_n^{(\alpha)}(x)$ are associated Laguerre polynomials and
$U(a,b,x)$ represents confluent hypergeometric (Kummer) function.
Multiplication by a finite power $x^\alpha$ merely shifts all
parameters $a_1,\ldots,a_n,b_1,\ldots,b_m$ by constant $\alpha$.
Most integrals encountered can be recast as those of Mellin type:
\begin{equation}
  \int_0^\infty 
  G_{p,q}^{m,n} \biggl( \frac{x}{t} \bigg| \stack{\vec{a}}{\vec{b}} \biggr)
  G_{r,s}^{k,\ell} \biggl( t \bigg| \stack{\vec{c}}{\vec{d}} \biggr)
  \frac{\dd t}{t} =
  G_{p+r,q+s}^{m+k,n+\ell}
  \biggl( x \bigg| \stack{\vec{a},\vec{c}}{\vec{b},\vec{d}} \biggr),
\end{equation}
where vectors are concatenated as follows: $\vec{a},\vec{c}$ consists
of $a_1,\ldots,a_n$ and $c_1,\ldots,c_\ell$ followed by
 $a_{n+1},\ldots,a_p$ and $c_{\ell+1},\ldots,c_r$ (similarly for
 $\vec{b}$ and $\vec{d}$). This ensures that zeros and poles in the
Mellin transform of G-functions are grouped together.

\smallskip
\textbf{Mean-Field Analysis.}
Self-consistency equation for the `magnetisation' vector defined in
eqn.~\eqref{m} of the main text can be written as follows:
\begin{equation}
  \mathbf{m} = \frac{1}{N}\sum_i \boldsymbol\xi_i 
  \langle \sigma_i^z \rangle
  = \frac{1}{N}\sum_i
  \frac{\boldsymbol\xi_i(\boldsymbol\xi_i \mathbf m)}
  {\sqrt{\Gamma^2+(\boldsymbol\xi_i \mathbf m)^2}},
\end{equation}
where the individual spins polarise along the magnetic field with
the components $\Gamma$ and $h_i = \boldsymbol\xi_i \mathbf m$ 
in the transverse and longitudinal directions respectively.
Replacing sum over spins with disorder average, one obtains
non-trivial solution to self-consistency equations for $\Gamma<1$.
For the bimodal distribution, the solutions are $(\pm\sqrt{1-\Gamma^2},0,\ldots,0)$, 
$(0,\pm\sqrt{1-\Gamma^2},\ldots,0)$, etc. Other spurious solutions
appear for smaller $\Gamma$, but they do not become stable for finite $p$.
For Gaussian distribution, the magnetisation vector can have arbitrary
direction while the magnitude $m_\Gamma$ satisfies
\begin{equation}
  1 = \int \tfrac{\xi^2}{\sqrt{\Gamma^2+\xi^2m_\Gamma^2}}
  \tfrac{1}{\sqrt{2\pi}} \ee^{-\xi^2/2} \dd\xi
  = \tfrac{1}{\sqrt{2}m_\Gamma} 
  U\Bigl(\tfrac{1}{2},0;\tfrac{\Gamma^2}{2m_\Gamma^2}\Bigr),  
  \label{mG}
\end{equation}
where the integral on the right hand side had been converted to Mellin
type, $\int_0^\infty G_{1,1}^{1,1} \bigl(\frac{x}{t} \big| \stack{1/2}{0} \bigr) 
G_{0,1}^{1,0} \bigl( t \big | \stack{-}{1} \bigr) \frac{\dd t}{t}$
by substituting $t=\xi^2/2$ and $x=\Gamma^2/(2m_\Gamma^2)$. 
Alternatively, it can be expressed in terms of modified Bessel
functions.

Low-energy spectrum of Hopfield model can be obtained by examining the
partition function at finite temperature. It can be written as a sum
over paths $[s_i(t)]$ alternating between values $s_i(t) = \pm 1$ with
periodic boundary conditions $s_i(\beta)=s_i(0)$:
\begin{equation}
  Z \sim \sum_{[\{s_i(t)\}]} \ee^{\frac{1}{2}\int_0^\beta J_{ik}s_i(t)s_k(t)\dd t
    +\sum_i \mathcal{K}_\Gamma[s_i(t)]}.
\end{equation}
The `kinetic' term $\mathcal{K}[s(t)] = \text{(\# of kinks)} \times
\frac{1}{2} \ln \coth \Gamma \Delta t$, where $\Delta t$ is the
discretisation chosen (limit $\Delta t \to 0$ will be taken eventually).
This term completely suppresses the dynamics in the limit $\Gamma=0$.

With $J_{ik}=(1/N)\sum_{\mu} \xi_i^{(\mu)} \xi_k^{(\mu)}$, two-body
interactions are decoupled via Hubbard-Stratonovich transformation:
Using the identity $\ee^{\frac{1}{2N}\left(\sum_i\boldsymbol\xi_i s_i\right)} \propto
\int \dd \mathbf m \ee^{-\frac{N}{2}\mathbf m^2+\sum_i\boldsymbol\xi_i \mathbf m s_i}$,
the partition function is recast as a path integral,
\begin{equation}
  Z \sim \int [\dd\mathbf{m}(t)] \ee^{
    -\frac{N}{2}\int_0^\beta\mathbf{m}^2(t)\dd t
    +\sum_i \ln Z_i},
  \label{Zpi}
\end{equation}
where
\begin{equation}
  Z_i = \sum_{[s(t)]} \ee^{\int_0^\beta h_i(t) s_i(t) \dd t + \mathcal K_\Gamma[s(t)]}
  = \Tr \mathcal T \ee^{-\int_0^\beta \hat H_i(t) \dd t}
  \label{Zi}
\end{equation}
is a single-site partition function for a spin in a `time-dependent'
longitudinal field $h_i(t) = \boldsymbol\xi_i \mathbf m(t)$. It has
been rewritten with the aid of time-ordering operator $\mathcal T$.
The single-site Hamiltonian $\hat H_i(t) = - h_i(t) \hat \sigma^z -
\Gamma \hat \sigma^x$.

The dominant contribution to the path integral is given by stationary
paths, $\mathbf{m}(t) \equiv \mathbf{m}$. To leading order
In this case $Z \sim \int \dd\mathbf{m} \ee^{-N \beta \mathcal{V}(\mathbf{m})}$,
where the effective potential is
\begin{equation}
  \mathcal{V}(\mathbf{m}) = \tfrac{\mathbf{m}^2}{2} - 
  \Bigl<\sqrt{\Gamma^2+(\boldsymbol\xi\mathbf{m})^2}\Bigr>_{\boldsymbol\xi}
  = \tfrac{\mathbf{m}^2}{2} - \sqrt{2} |\mathbf{m}|
  U \Bigl( -\tfrac{1}{2}; 0; \tfrac{\Gamma^2}{2\mathbf{m}^2} \Bigr).
\end{equation}
The saddle-point value $m_\Gamma=|\mathbf{m}|$ that minimises the
potential for $\Gamma=0$ is given by a solution of eqn.~\eqref{mG}.

Considering higher-order corrections, the dominant contribution comes
from paths where the magnitude of magnetisation fluctuates around the
mean-field value and the angle is a slow function of time:
$\mathbf{m}(t) \approx m_\Gamma \bigl( \begin{smallmatrix}
  -\sin \vartheta(t) \\ \cos \vartheta(t) \end{smallmatrix} \bigr)$.
Since the local field $h_i(t)$ is slow-varying, it is convenient to
evaluate the partition function \eqref{Zi} in the adiabatic basis that
diagonalises the instantaneous 2-level Hamiltonian:
$\hat H_i(t) = \ee^{\ii \hat V_i(t)} \hat E_i(t) \ee^{-\ii \hat V_i(t)}$
where
\begin{equation}
  \hat E_i(t) = -\sqrt{\Gamma^2+h_i^2(t)} \, \hat\sigma^z
  \text{ and }
  \hat V_i(t) = \frac{1}{2} \arccot \frac{h_i(t)}{\Gamma} \hat\sigma^y.
\end{equation}
Using this factorisation, the time-ordered product
$\Tr \bigl( \cdots \ee^{\ii \hat V_i(t+\Delta t)} \ee^{-E_i(t+\Delta t)} \ee^{-\ii \hat V_i(t+\Delta t)}
\ee^{\ii \hat V_i(t)} \ee^{-E_i(t)} \ee^{-\ii \hat V_i(t)} \cdots \bigr)$
can be rewritten in the limit $\Delta t \to 0$ as
\begin{equation}
  Z_i = \Tr \mathcal{T} \ee^{-\int_0^\beta \left( \hat E(t)+
      \ii \frac{\partial}{\partial t} \hat V(t) \right) \dd t}
\end{equation}
The non-adiabatic term can be simply written as $\ii \frac{\partial}{\partial t} \hat V(t)$ 
since all $\hat V(t) \propto \hat\sigma^y$ are commuting (should be 
$-U^\dag\frac{\partial}{\partial t}U$ more generally) and acts as a perturbation.
The lower-energy level, $E_\downarrow(t) = -\sqrt{\Gamma^2+h_i^2(t)}$,
makes a dominant contribution to the partition function in the limit
$\beta \to \infty$. Including second-order correction from the
perturbation theory, $|\partial V_{\downarrow\uparrow}/\partial
t|^2/(E_\downarrow-E_\uparrow)$, one obtains:
\begin{equation}
  \ln Z_i \approx \int_0^\beta \biggl(
  \sqrt{\Gamma^2+h_i^2} -
  \frac{\Gamma^2 (\partial h_i/\partial t)^2}
  {8 (\Gamma^2+h_i^2)^{5/2}} \biggr) \dd t.
  \label{Ziint}
\end{equation}
As the sum over all sites is performed, the first term in the
integrand gives rise to a finite-size correction to the effective
potential, which depends on angle $\vartheta$ and a particular
realisation of disorder
\begin{equation}
  V_\Gamma(\vartheta) = -\sum_i \sqrt{\Gamma^2+
    \xi_i^2m_\Gamma^2\sin^2(\vartheta-\theta_i)}+
  N\Bigl< \sqrt{\Gamma^2+\xi^2 m_\Gamma^2} \Bigr>_\xi.
  \label{VGdef}
\end{equation}
Vector-valued $\boldsymbol\xi_i$ has been parametrised
here as $\xi_i \bigl( \begin{smallmatrix} \cos \vartheta(t) \\ 
\sin \vartheta(t) \end{smallmatrix} \bigr)$. Offsetting average in
eqn.~\eqref{VGdef} above has been made with respect to Gaussian
distribution of the projection of $\boldsymbol \xi$ onto
$\mathbf m$ (its value is independent of the direction).
The second term in the integrand of eqn.~\eqref{Ziint} contributes 
\begin{equation}
  \int_0^\beta \frac{1}{2} \underbrace{\sum_i 
  \frac{\Gamma^2 \xi_i^2 m_\Gamma^2 \cos^2(\vartheta-\theta_i)}
  {4\left(\Gamma^2 + \xi_i^2 m_\Gamma^2 \sin^2(\vartheta-\theta_i)\right)^{5/2}}}_M
  \biggl( \frac{\dd\vartheta}{\dd t} \biggr)^2 \dd t. 
\end{equation}
The `effective mass' $M$ can be replaced by its average value
\begin{equation}
  M = \frac{N m_\Gamma}{4 \sqrt{2} \Gamma^2} 
  U \Bigl( \tfrac{1}{2}, -1; \tfrac{\Gamma^2}{2 m_\Gamma^2} \Bigr),
\end{equation}
which approaches $N\big/\bigl(4\sqrt{\pi} \Gamma^2\bigr)$ in the limit
$\Gamma \to 0$. The dynamics becomes more classical as the effective
mass diverges. Path integral \eqref{Zpi} is then rewritten as
\begin{equation}
  Z \sim \ee^{- N \beta \mathcal{V}(m_\Gamma)}
  \int [\dd \vartheta(t)] \ee^{-\int_0^\beta \left(
      \frac{M}{2}\left(\frac{\dd \vartheta}{\dd t}\right)^2
      + V_\Gamma(\vartheta) \right) \dd t}.
\end{equation}
The low-lying energy levels of the many-body problem are thus
equivalent to those of a quantum mechanical particle on a ring.

The approximations made above break down for any finite $N$ when $\Gamma$ is
sufficiently small. To avoid this, and to be able to make more
quantitative predictions about the behavior of random potential \eqref{VGdef},
it is necessary to study the continuous limit $N \to \infty$ directly.

\smallskip
\textbf{Continuous Random Potential.}
Observe that the central limit theorem can be applied to the entire
two-dimensional (in $\Gamma$ and $\vartheta$) process
$V_\Gamma(\vartheta)$. This Gaussian process can be decorrelated by
writing it as an infinite function series 
$V_\Gamma(\vartheta)=\sum_{n,m} \zeta_{nm} f_{nm}(\Gamma,\vartheta)$
involving independent Gaussian variables $\zeta_{nm}$.
Basis functions are determined from the covariance 
$\langle V_\Gamma(\vartheta)V_{\Gamma'}(\vartheta') \rangle$.
Since this correlation depends only on the angle difference
$\vartheta-\vartheta'$, it is natural to use
$f_{nm}(\Gamma,\vartheta) = \tilde f_{nm}(\Gamma)\,\ee^{\ii m \vartheta}$.

Although the best truncated approximation is obtained by using 
Karhunen-Lo\`eve basis, there is no requirement that
$\tilde f_{nm}(\Gamma)$ be orthogonal. In fact, since covariance
involves averaging over distribution $\mathcal P(\xi)=\xi\,\ee^{-\xi^2/2}$, 
it is quite convenient to use a set of polynomials that are
orthogonal with said weight. Using a set of white noise processes
\begin{equation}
  \bigl< \zeta_n(\vartheta) \, \zeta_{n'}(\vartheta') \bigr> = 
  \delta_{nn'} \, \delta(\vartheta-\vartheta')
  \text{ for } \vartheta,\vartheta' \in [-\pi/2; \pi/2],
\end{equation}
in lieu of discrete random variables, the series expansion becomes
\begin{equation}
  V_\Gamma(\vartheta) = \sqrt{N} \sum_{n=0}^\infty \int_{-\pi/2}^{\pi/2}
  f_\Gamma^{(n)} (\vartheta-\theta) \, \zeta_n(\theta) \,\dd \theta,
  \label{VGfn}
\end{equation}
where a factor $\sqrt{N}$ had been introduced for convenience. 
The convolution kernels should be chosen so as to match the covariance.
One suitable choice is

\begin{equation}
  f_{\Gamma}^{(n)}(\vartheta) = A_n \!\int_0^\infty\!
  \sqrt{\Gamma^2 + m_\Gamma^2 \xi^2 \sin^2 \vartheta} \,\xi^{\alpha + 1}
  \ee^{-\xi^2/2} L_n^{(\alpha)}\!\Bigl(\tfrac{\xi^2}{2}\Bigr) \dd\xi.
  \label{fn}
\end{equation}
Since associated Laguerre polynomials form a complete orthogonal
(with respect to weight $t^\alpha \ee^{-t}$ with $t=\xi^2/2$) set of basis
function, it follows that
\begin{equation}
  \sum_{n=0}^\infty f_\Gamma^{(n)}(\vartheta)
  L_n^{(\alpha)}\!\Bigl(\tfrac{\xi^2}{2}\Bigr) =
  \frac{2^\alpha\boldsymbol\Gamma(n+\alpha+1)A_n}{n!\xi^\alpha}
  \sqrt{\Gamma^2+m_\Gamma^2 \xi^2 \sin^2\vartheta}.
  \label{fnLn}
\end{equation}
Representing $V_\Gamma(\vartheta)$ using ansatz of eqn.~\eqref{VGfn} 
above one can write the expression for the covariance matrix. 
Using identities \eqref{fn} and \eqref{fnLn}, it is transformed as follows:
\begin{multline}
  \bigl< V_\Gamma(\vartheta) V_{\Gamma'}(\vartheta') \bigr> =
  \sqrt{N} \sum_{n=0}^\infty \int_{-\frac{\pi}{2}}^{\frac{\pi}{2}} \!
  f_\Gamma^{(n)}\!(\vartheta-\theta)
  \bigl<\zeta_n(\theta) V_{\Gamma'}(\vartheta')\bigr> \dd\theta\\
  \shoveleft{
    = N A_n \sum_{n=0}^\infty \int_{-\frac{\pi}{2}}^{\frac{\pi}{2}}\!
    f_\Gamma^{(n)}\!(\vartheta-\theta) \times} \\
    \int_0^\infty\!\sqrt{{\Gamma'}^2+m_{\Gamma'}^2\xi^2\sin^2(\vartheta'-\theta)}
    \,\xi^{\alpha+1} \ee^{-\xi^2/2} L_n^{(\alpha)}\!\Bigl(\tfrac{\xi^2}{2}\Bigr)
    \,\dd\xi \, \dd\theta \\[2ex]
  \shoveleft{
    = \frac{2^\alpha \boldsymbol\Gamma(n+\alpha+1) N A_n^2}{n!} 
    \int_{-\frac{\pi}{2}}^{\frac{\pi}{2}} \! \int_0^\infty \!
    \sqrt{\Gamma^2+m_\Gamma^2\xi^2\sin^2(\vartheta-\theta)}\times}\\
  \sqrt{{\Gamma'}^2+m_{\Gamma'}^2\xi^2\sin^2(\vartheta'-\theta)}
  \,\xi\,\ee^{-\xi^2/2} \dd\xi\, \dd\theta.
\end{multline}
This matches the correct value obtained by replacing sum over sites
by disorder averages and determines the normalisation constant:
\begin{equation}
  A_n = \sqrt{\frac{n!}{2^\alpha \pi \boldsymbol\Gamma(n+\alpha+1)}}\,.
\end{equation}

Notice that the value of $\alpha$ can be chosen freely: as mentioned
above, the decomposition is not unique. The most convenient choice,
$\alpha=1$, leads to the vanishing of all convolution kernels
with $n \geqslant 1$ in the limit $\Gamma=0$. Introducing the rescaled
potential via $V_0(\vartheta)=\bigl(2\sqrt{N}/\pi\bigr) \chi(\vartheta)$,
it is straightforward to verify that
\begin{equation}
  \chi(\vartheta) = \int_{-\frac{\pi}{2}}^{\frac{\pi}{2}}\!
  \tfrac{1}{2} \bigl|\sin(\vartheta-\theta)\bigr|\,
  \zeta_0(\theta) \, \dd \theta
\end{equation}
satisfies the following stochastic equation:
\begin{equation}
  \frac{\dd^2 \chi}{\dd \vartheta^2} + \chi = \zeta_0(\vartheta).
  \label{chizeta}
\end{equation}
The solution is determined uniquely by enforcing periodic boundary
conditions: $\chi(-\pi/2)=\chi(\pi/2)$ and $\chi'(-\pi/2)=\chi'(\pi/2)$.

Now considering a finite $\Gamma>0$, it becomes convenient to reexpress
$n=0$ term in the functional series as a convolution with $\chi(\vartheta)$
instead. This can be accomplished by substituting the left hand side
of \eqref{chizeta} into eqn.~\eqref{VGfn} and performing integration by
parts. Other terms can also be more conveniently expressed by
rewriting white-noise processes in terms of continuous processes:
realisations of periodic brownian motion
\begin{equation}
  \frac{\dd\eta_n}{\dd\vartheta} + \bar \zeta_n = \zeta_n(\vartheta).
\end{equation}
Here $\bar \zeta_n = \frac{1}{\pi} \int_{-\pi/2}^{\pi/2}\!
\zeta_n(\vartheta)\, \dd\vartheta$ is added to ensure periodicity:
$\eta_n(-\pi/2)=\eta_n(\pi/2)$.

Now, the random potential is rewritten as
\begin{equation}
  V_\Gamma(\vartheta) = \sqrt{N} 
  \,\frac{m_\Gamma}{\sqrt{\pi/2}}
  \biggl(
  (\mathfrak{f}_\Gamma \star \chi)(\vartheta) + 
  \sum_{n=1}^\infty \bigl(\mathfrak{g}_\Gamma^{(n)} \!\star \eta_n\bigr)(\vartheta)
  + \const \biggr).
\end{equation}
The last term is angle-independent random variable that absorbs
$\{\bar \zeta_n \}$. Being a constant offset, it cannot affect the physics;
and, thus, will be neglected. Expressions for kernels 
$\mathfrak f_\Gamma(\vartheta)$ and
$\mathfrak g_\Gamma^{(n)}$ are obtained by performing integration by parts in
variable $\vartheta$:
\begin{align}
  \mathfrak{f}_\Gamma(\vartheta) &= \int_0^\infty 
  \frac{\widetilde\Gamma^2\bigl(\widetilde\Gamma^2+\xi^2\bigr)\xi^2}
  {2\bigl(\widetilde\Gamma^2+\xi^2\sin^2\vartheta\bigr)^{3/2}}
  \ee^{-\xi^2/2} \dd \xi, \\
  \mathfrak{g}_\Gamma^{(n)}\!(\vartheta) &= 
  \frac{\sin\vartheta\cos\vartheta}{2\sqrt{n+1}}
  \int_0^\infty \frac{\xi^4 L_n^{(1)}\!\bigl(\tfrac{\xi^2}{2}\bigr)}
  {\sqrt{\widetilde\Gamma^2+\xi^2\sin^2\vartheta}} \ee^{-\xi^2/2} \dd\xi.
\end{align}
where the rescaled transverse field $\widetilde\Gamma=\Gamma/m_\Gamma$
has been introduced for convenience. Using substitutions $t=\xi^2/2$
and $x=\widetilde \Gamma^2/(2 \sin^2 \vartheta)$, the integrals above
can be converted to Mellin form and evaluated in terms of special functions:
\begin{align}
  \mathfrak{f}_\Gamma(\vartheta) &= 
  \frac{\sqrt{\pi}\,\widetilde\Gamma^2}{8\sin^3\vartheta}
  \left[
  3 U\biggl(\frac{3}{2},0;\frac{\widetilde\Gamma^2}{2\sin^2\vartheta}\biggr)
    +\widetilde\Gamma^2
    U\biggl(\frac{3}{2},1;\frac{\widetilde\Gamma^2}{2\sin^2\vartheta}\biggr)
  \right], \label{fG} \\
  \mathfrak{g}_\Gamma^{(n)}\!(\vartheta) &= 
  \frac{\sqrt{\pi}\,\cos\vartheta}{\sqrt{n+1}\,n!}
  G_{2,3}^{2,2}\biggl( \frac{\widetilde\Gamma^2}{2\sin^2\vartheta}
  \bigg| \stack{1-n,\frac{1}{2}}{1,2;0} \biggr).  \label{gG}
\end{align}

Tunneling gaps are associated with global bifurcations of
$V_\Gamma(\vartheta)$ which occur mostly in the limit $\Gamma \ll 1$.
Moreover, it is this limit that is governed by universal scaling laws
and thus will be investigated in greater detail. The kernels above
act on scales $\vartheta \sim \Gamma$. For small $\Gamma$ it is
permissible to drop the second term in eqn.~\eqref{fG} and to replace
$\sin \vartheta \approx \vartheta$ and $\cos \vartheta \approx 1$
everywhere. It will be seen that any bifurcations also take place
for $\vartheta \sim \Gamma$, hence the periodicity of the potential
$V_\Gamma(\vartheta)$ becomes irrelevant.

\smallskip
\textbf{Extremal Statistics.}
The distribution of bottlenecks in the limit $\Gamma \ll 1$ is related
to the properties of the classical potential $\chi(\vartheta)$ in a
proximity of its global minimum $\vartheta_0^\ast$. It will be governed
by different statistics, derived below, by virtue of this extremality condition.

On short scales, the linear term on the left hand side of
eqn.~\eqref{chizeta} can be neglected in comparison with the large stochastic
term $\zeta_0(\vartheta)$. It will be seen self-consistently that
the relevant range is $|\vartheta-\vartheta_0^\ast|=O(\Gamma)$.
The stochastic equation describes a free particle in 1D subjected to
random force. Specifying the `coordinate' $\chi$ together with the
`velocity' $\upsilon=\dd\chi/\dd\vartheta$ for some $\vartheta$
determines their distribution either in the `future' ($\vartheta'>\vartheta$)
or in the `past' ($\vartheta'<\vartheta$), where $\vartheta$ plays the
role of time. This can be described by the following PDE:

\begin{equation}
  \frac{\partial p}{\partial \vartheta} 
  + \upsilon \frac{\partial p}{\partial \chi} -
  \frac{1}{2} \frac{\partial^2 p}{\partial \upsilon^2} = 0,
  \label{pdep}
\end{equation}
where $p(\vartheta;\chi,\upsilon)$ is the probability density. 
It will be convenient to perform a shift so that a global minimum
coincides with the origin: $\vartheta_0^\ast=0$ and $\chi(0)=0$.
Another condition is necessary to ensure that $\chi(\vartheta)>0$
for $\vartheta \neq 0$. Considering only $\vartheta>0$ for
concreteness, one may introduce an absorbing boundary via
boundary condition
\begin{equation}
  \lim_{\chi \to +0} p(\vartheta;\chi,\upsilon)=0
  \quad \text{for } \upsilon>0.
  \label{bc}
\end{equation}
Notice that the probability need not vanish as $\chi\to+0$ for
negative velocities $\upsilon<0$. The probability that a random path
does not hit this boundary decays as follows:
\begin{equation}
  \frac{\partial}{\partial\vartheta} 
  \int p(\vartheta;\chi,\upsilon)\, \dd \chi\, \dd \upsilon
  = \int_{-\infty}^0 \upsilon\, p(\vartheta;+0,\upsilon)\, \dd\upsilon.
\end{equation}
Since the random process is conditioned on the fact that it starts at
a global minimum, it is convenient to `renormalise' the probability so
that it becomes a conserved quantity once again. Consider a
probability that $\chi$ stays positive at least until some distant
horizon $\Theta$,
\begin{equation}
  q(\vartheta;\chi,\upsilon) \propto p(\vartheta;\chi,\upsilon)
  \underbrace{
  \int_{\Theta>0} P(\Theta;\Chi,\Upsilon|\vartheta;\chi,\upsilon)\,
  \dd\Chi\, \dd\Upsilon}_{P_\Theta(\chi,\upsilon,\vartheta)},
\end{equation}
with an appropriate normalisation factor. The second factor,
$P_\Theta(\vartheta;\chi,\upsilon)$ represents the survival probability.
It is written as an integral of the Green's function associated with
eqn.~\eqref{pdep}. Time-reversal symmetry implies
\begin{equation}
  P(\Theta;\Chi,\Upsilon|\vartheta;\chi,\upsilon) = 
  P(\vartheta;\chi,-\upsilon|\Theta;\Chi,-\Upsilon).
\end{equation}
As a result $P_\Theta(\vartheta;\chi,\upsilon)$ satisfies a PDE obtained
from eqn.~\eqref{pdep} by changing the sign of 
$\frac{\partial p}{\partial\vartheta}$ and $\upsilon
\frac{\partial p}{\partial\chi}$. With this in mind,
the equation for renormalised probability becomes
\begin{equation}
  \frac{\partial q}{\partial \vartheta}
  + \upsilon \frac{\partial q}{\partial \chi}
  + \frac{\partial}{\partial \upsilon} \biggl(
  \frac{1}{P_\Theta} \frac{\partial P_\Theta}{\partial \upsilon} q \biggr)
  - \frac{1}{2} \frac{\partial^2 q}{\partial \upsilon^2} = 0.
  \label{pdeq}
\end{equation}
In comparison with eqn.~\eqref{pdep}, it adds a drift $\propto \partial
(\log P_\Theta)/\partial \upsilon$ in addition to the diffusion of
`velocity' $\upsilon$.

In the limit of large $\Theta$, the `survival' probability
$P_\Theta(\chi,\upsilon,\vartheta)$ corresponds --- up to time-reversal
($\upsilon \to -\upsilon$) --- to the asymptotic solution of \eqref{pdep},
\begin{equation}
  p(\chi,\upsilon,\vartheta) \sim 
  A \frac{p_{\ast}(\chi,\upsilon)}{\vartheta^{\alpha}}.
\end{equation}
Probability density always converges to this form as $\vartheta \to \infty$,
with dependence on the initial conditions only via the prefactor $A$.
Here $p_\ast(\chi,\upsilon)$ can be thought of as a `stationary'
solution: it satisfies eqn.~\eqref{pdep} (with $\partial/\partial\vartheta$ 
term dropped) and boundary condition \eqref{bc}.
Eqn.~\eqref{pdep} is invariant with respect to rescaling
$(\vartheta,\chi,\upsilon) \mapsto 
(\ell\vartheta,\ell^{3/2}\chi,\ell^{1/2}\upsilon)$.
For this reason, the asymptotic solution should be sought in the form
\begin{equation}
  p_\ast(\chi,\upsilon) = \chi^{2\alpha/3} \,
  p_\ast\bigl(\upsilon/\chi^{1/3}\bigr).
  \label{past}
\end{equation}

It is possible to simplify equations even further by introducing new
`dimensionless' variables. Defining new `time' variable via $\dd \tau =
\dd \vartheta\big/\chi^{2/3}$, it is now a triple $(\vartheta,\chi,\upsilon)$
that becomes new Markov process. This modification merely adds
$\chi^{-2/3}(\partial q/\partial \tau) + \frac{2}{3}(\upsilon/\chi)q$
to the left hand side of eqn.~\eqref{pdeq}. The term $\propto q$
reflects renormalisation of probability density as a result of
non-linear transformation.

Furthermore, it is convenient to introduce new `dimensionless' velocity
$\nu=\upsilon/\chi^{1/3}$ and the `logarithmic' coordinate 
$\mu=\ln\chi$. The new probability density, appropriately renormalised,
reads
\begin{equation}
  \bar q(\tau;\vartheta,\mu,\nu) = 
  \ee^{4\mu/3} q\bigl(\tau;\vartheta,\ee^\mu,\nu\,\ee^{\mu/3}\bigr).
\end{equation}
This is substituted into eqn.~\eqref{pdeq} which has been modified as described in
the previous paragraph. Replacing $\partial/\partial\chi \mapsto
\ee^{-\mu}\bigl(\partial/\partial\mu-\frac{1}{3}\nu\,\partial/\partial\nu\bigr)$
and $\partial/\partial\upsilon \mapsto \ee^{-\mu/3}\partial/\partial\nu$,
the following PDE is obtained:
\begin{equation}
  \frac{\partial \bar q}{\partial\tau}
  + \ee^{2\mu/3} \frac{\partial \bar q}{\partial\vartheta}
  + \nu \frac{\partial \bar q}{\partial\mu}
  - \frac{\partial}{\partial\nu} \biggl(
  \frac{\partial \mathcal U}{\partial \nu} 
  \bar q \biggr)
  - \frac{1}{2} \frac{\partial^2 \bar q}{\partial \nu^2} = 0,
  \label{pdeqbar}
\end{equation}
where the non-linear potential is written, using eqn.~\eqref{past}, as
\begin{equation}
  \mathcal U(\nu) \sim \frac{\nu^3}{9} - \ln p_\ast(-\nu).
\end{equation}
When $p(\tau;\vartheta,\mu,\nu)$ is marginalised over $\vartheta$, $\mu$,
the standard Fokker-Planck equation is obtained. It describes a
stochastic process
\begin{equation}
  \frac{\dd \nu}{\dd \tau} = -\mathcal U'(\nu) \sgn \tau +\zeta(\tau),
  \label{dnudt}
\end{equation}
where $\zeta(\tau)$ is the white-noise random force. This stochastic
equation can be integrated both forward ($\tau>0$) and backwards
($\tau<0$) starting from $\nu(0)$ drawn from the equilibrium distribution
$\rho(\nu) \propto \ee^{-2\mathcal U(\nu)}$.

To find the shape of $\mathcal U(\nu)$ and $\rho(\nu)$, observe
that $p_\ast\bigl(\upsilon\big/\chi^{1/3}\bigr)$ satisfies a stationary version
of eqn.~\eqref{pdep}, which is reduced to ordinary differential
equation. Writing $z=\nu^3/9$, 
\begin{equation}
  z \psi'' + \frac{2}{3} \psi' + \Bigl( \frac{2+4\alpha}{3} - z \Bigr) \psi = 0,
  \label{ode}
\end{equation}
where $\psi(z)=\sqrt{\rho(z)}$. Disallowing exponentially increasing
solutions, $\psi(z)$ is sought in the form $e^{\mp z} f_\pm(z)$ for
$z>0$ and $z<0$ respectively.
This substitution transforms eqn.~\eqref{ode} to hypergeometric form,
giving the two branches as $f_+(z) \propto U \left(
-\frac{2\alpha}{3}; \frac{2}{3}; 2z \right)$ and 
$f_-(z) \propto U\left( \frac{2}{3}+\frac{2\alpha}{3}; \frac{2}{3};
-2z \right)$. Matching logarithmic derivatives
\begin{equation}
  \frac{\psi'(+0)}{\psi(+0)} = 
  - A \frac{\boldsymbol\Gamma\!\left(\frac{1}{3}-\frac{2\alpha}{3}\right)}
  {\boldsymbol\Gamma\!\left(-\frac{2\alpha}{3}\right)}
  \text{ and }
  \frac{\psi'(-0)}{\psi(-0)} =
  A \frac{\boldsymbol\Gamma\!\left(1+\frac{2\alpha}{3}\right)}
  {\boldsymbol\Gamma\!\left(\frac{2}{3}+\frac{2\alpha}{3}\right)}
\end{equation}
[where the common factor is $A=3^{1/3}
\boldsymbol\Gamma\!\left(\tfrac{2}{3}\right)\!/
\boldsymbol\Gamma\!\left(\tfrac{1}{3}\right)$] requires that
\begin{equation}
  \alpha = \frac{1}{4} + \frac{3}{2}n,
\end{equation}
as verified using of Euler's identity $\boldsymbol\Gamma(x)
\boldsymbol\Gamma(1-x) = \pi / \sin(\pi x)$. Of this infinite set,
the smallest positive value $\alpha=1/4$ has to be chosen.
The analytical expression that defines the potential $\mathcal
U(\nu)$ can be simplified as follows:
\begin{equation}
  \psi(\nu) \propto \ee^{-\mathcal U(\nu)} = 
  x \Ai(x^2) - \Ai'(x^2),
  \quad \text{where } x=\nu\big/\sqrt[3]{6}.
\end{equation}

Since eqn.~\eqref{pdeqbar} is first order in $\vartheta$, $\mu$, these are
completely deterministic, given a particular realisation of stochastic
process \eqref{dnudt}:
\begin{align}
  \frac{\dd \mu}{\dd \tau} &= \nu(\tau), \\
  \frac{\dd \vartheta}{\dd \tau} &= \chi^{2/3}(\tau),
\end{align}
with boundary conditions $\chi(-\infty)=\vartheta(-\infty)=0$. 
This procedure defines the process $\chi(\vartheta)$ parametrically.
So far, a positive branch ($\vartheta>0$) had been considered.
Negative branch ($\vartheta<0$) is obtained similarly; positive and
negative branches are uncorrelated.

\end{footnotesize}


\begin{thebibliography}{99}

\bibitem{qcbook} 
  Nielsen, M. \& Chuang, I.
  {\it Quantum Computation and Quantum Information},
  Cambridge University Press, Cambridge, MA,
  10th ed. (2011).

\bibitem{shor}
  Shor, P.~W.
  Algorithms for quantum computation: disctete logarithms and factoring.
  {\it Proc. 35th IEEE Symp. FOCS,} 124--134 (1994).
  
\bibitem{np}
  Garey M.~R. \& Johnson D.~S.
  {\it Computers and Intractability: A Guide to the Theory of NP-Completeness.}
  W.~H.~Freeman, New York, NY (1979).

\bibitem{nishimori}
  Kadowaki, T. \& Nishimori, H.
  Quantum Annealing in the Transverse Ising Model.
  {\it Phys. Rev. E} {\bf 58,} 5355 (1998).

\bibitem{farhiX}
  Farhi, E., Goldstone, J., Gutmann, S., \& Sipser M.
  Quantum Computation by Adiabatic Evolution.
  Preprint at http://arxiv.org/abs/quant-ph/0001106 (2000).

\bibitem{rmp}
  Das, A. \& Chakrabarti, B.~K.
  Quantum annealing and analog quantum computation.
  {\it Rev. Mod. Phys.} {\bf 80,} 1061--1081 (2008).

\bibitem{vandam}
  van~Dam, W., Mosca, M. \& Vazirani, U.
  How powerful is adiabatic quantum computation?
  {\it Proc. 42nd IEEE Symp. FOCS,} 279--287 (2001).

\bibitem{brooke}
  Brooke, J., Bitko, D., Rosenbaum, F.T. \& Aeppli, G.
  Quantum annealing of a disordered magnet.
  {\it Science} {\bf 284,} 779--781 (1999).

\bibitem{santoro}
  Santoro, G., Marto\v{n}\'{a}k, R., Tosatti, E. \& Car, R.
  Theory of quantum annealing of spin glasses.
  {\it Science} {\bf 295,} 2427--2430 (2002).

\bibitem{heim}
  Heim, B., R{\o}nnow, T.~F., Isakov, S.~V. \& Troyer, M.
  Quantum versus classical annealing of Ising spin glasses.
  {\it Science} {\bf 348,} 215--217 (2015).

\bibitem{farhiSci}
  Farhi, E., Goldstone, J., Gutmann, S., Lapan, J., Lundgren, A. \& Preda, D.
  A quantum adiabatic evolution algorithm applied to random instances
  of an NP-complete problem.
  {\it Science} {\bf 292,} 472--475 (2001).

\bibitem{young}
  Young, A.~P., Knysh, S., \& Smelyanskiy, V.~N.
  First order phase transition in the quantum adiabatic algorithm.
  {\it Phys. Rev. Lett} {\bf 104,} 020502 (2010).
  
\bibitem{boixo}
  Boixo, S. {\it et al.}
  Evidence for quantum annealing with more than one hundred qubits.
  {\it Nat. Phys.} {\bf 10,} 218--224 (2014).

\bibitem{ronnow}
  R{\o}nnow, T.~F. et al.
  Defining and detecting quantum speedup.
  {\it Science} {\bf 345,} 420--424 (2014).

\bibitem{boixo2}
  Boixo, S. {\it et al.}
  Computational role of multiqubit tunneling in a quantum annealer.
  {\it Nat. Comm.}  {\bf 7,} 10327 (2016).

\bibitem{katzgraber}
  Katzgraber, H.~G., Hamze, F., Zhu, Z., Ochoa, A.~J. \& Munos-Bauza, H.
  Seeking Quantum Speedup through Spin Glasses: The Good, the Bad, and
  the Ugly.
  {\it Phys. Rev. X} {\bf 5,} 031026 (2015).

\bibitem{zhu}
  Zhu, Z., Ochoa A.~J., Schnabel S., Hamze, F. \& Katzgraber, H.~G.
  Best-case performance of quantum annealers on native spin-glass
  benchmarks: How chaos can affect success probabilities.
  {\it Phys. Rev. A} {\bf 93,} 012317 (2016).

\bibitem{partitioning}
  Smelyanskiy, V.~N., von~Toussaint, U. \& Timucin, D.~A.
  Dynamics of quantum adiabatic evolution algorithm for number partitioning.
  Preprint at http://arxiv.org/abs/quant-ph/0202155 (2002).

\bibitem{qrem1}
  Goldschmidt, Y.~Y.
  {\it Phys. Rev. B} {\bf 41}, 4858(R) (1990).

\bibitem{qrem2}
  J\"org, T., Krz\k{a}ka{\l}a, F., Kurchan, J. \& Maggs, A.~C.
  Simple glass models and their quantum annealing.
  {\it Phys. Rev. Lett.} {\bf 101}, 147204 (2008).

\bibitem{kxor}
  J\"org, T., Krz\k{a}ka{\l}a, F., Semerjian, G. \& Zamponi, F.
  First-order transitions and the performance of quantum algorithms in
  random optimization problems.
  {\it Phys. Rev. Lett.} {\bf 104}, 207206 (2010).
  
\bibitem{fisher}
  Fisher, D.~S.
  Critical behavior of random transverse-field Ising spin chains.
  {\it Phys. Rev. B} {\bf 51,} 6411--6461 (1995).

\bibitem{sk1}
  Miller, J. \& Huse, D.
  Zero-temperature critical behavior of the infinite-range quantum Ising spin glass.
  {\it Phys. Rev. Lett.} {\bf 70,} 3147--3150 (1993).

\bibitem{sk2}
  Ye, J., Sachdev, S. \& Read, N.
  Solvable spin glass of quantum rotors.
  {\it Phys. Rev. Lett.} {\bf 70,} 4011--4014 (1993).

\bibitem{sk3}
  Read, N., Sachdev, S. \& Ye, J.
  Landau theory of quantum spin glasses of rotors and Ising spins.
  {\it Phys. Rev. B} {\bf 52,} 384--410 (1995).

\bibitem{altshuler}
  Altshuler, B., Krovi, H. \& Roland, J.
  Anderson localization casts clouds over adiabatic quantum optimization,
  {\it Proc. Nat. Acad. Sci. USA} {\bf 107,} 12446--12450 (2010).

\bibitem{farhigap}
  Farhi, E., Gosset, D., Hen, I., Sandvik, A.~W., Shor, P., Young, A.~P. \& Zamponi, F.
  The performance of the quantum adiabatic algorithm on random
  instances of two optimization problems on regular hypergraphs.
  {\it Phys. Rev. A} {\bf 86,} 052334 (2012).

\bibitem{comment}
  Knysh, S. \& Smelyanskiy, V.~N.
  On the relevance of avoided crossings away from quantum critical
  point to the complexity of quantum adiabatic algorithm.
  Preprint at http://arxiv.org/abs/1005.3011 (2010).

\bibitem{laumann}
  Laumann, C.~R., Moessner, R., Scardicchio, A. \& Sondhi, S.~L.
  Quantum annealing: the fastest route to quantum computation?
  {\it Eur. Phys. J. ST} {\bf 224,} 75--88 (2015).

\bibitem{tchpow}
  Krz\k{a}ka{\l}a, F. \& Martin, O.~C.
  Chaotic temperature dependence in a model of spin glasses.
  {\it Eur. Phys. J. B} {\bf 28,} 199--209 (2002). 

\bibitem{hopfield}
  Hopfield, J.
  Neural networks and physical systems with emergent collective
  computational abilities.
  {\it Proc. Nat. Acad. Sci.} {\bf 79,} 2554--2558 (1982).

\bibitem{qhm}
  Nishimori, H. \& Nonomura, Y.
  Quantum Effects in Neural Networks.
  {\it J. Phys. Soc. Japan} {\bf 65,} 3780--3796 (1996).

\bibitem{polyfirst}
  Laumann, C.~R., Moessner, R., Scardicchio, A. \& Sondhi S.~L.
  The quantum adiabatic algorithm and scaling of gaps at first order
  quantum phase transitions.
  {\it Phys. Rev. Lett.} {\bf 109,} 030502 (2012).

\bibitem{bovier}
  Bovier, A., van~Enter, A.~C.~D. \& Niederhauser, B.
  Stochastic symmetry-breaking in a gaussian Hopfield model.
  {\it J. Stat. Phys.} {\bf 95,} 181--213 (1999).

\bibitem{nnbook}
  Hertz, J., Krogh, A. \& Palmer, R.~G.
  {\it Introduction to the Theory of Neural Computation},
  Addison-Wesley, Boston, MA (1995).
 
\bibitem{sinai}
  Sinai, Y.~G.
  On the distribution of some functions of the integral of a random walk.
  {\it Theor. Math. Phys.} {\bf 90,} 219--241 (1992).

\bibitem{groeneboom}
  Groeneboom, P., Jongbloed, G. \& Wellner, J.~A.
  Integrated brownian motion, conditioned to be positive.
  {\it Ann. Prob.} {\bf 27,} 1283--1303 (1999).

\bibitem{chaos}
  Bray, A.~J. \& Moore, M.~A.
  Chaotic nature of the spin-glass phase.
  {\it Phys. Rev. Lett.} {\bf 58,} 57--60 (1987).

\bibitem{tchsk1}
  Mulet, R., Pagnani, A. \& Parisi, G.
  Against temperature chaos in naive Thouless-Anderson-Palmer equations.
  {\it Phys. Rev. B} {\bf 63,} 184438 (2001).

\bibitem{rizzo}
  Rizzo, T. \& Crisanti, A.
  Chaos in temperature in the Sherrington-Kirkpatrick model.
  {\it Phys. Rev. Lett.} {\bf 90,} 137201 (2003).

\bibitem{tchsk2}
  Billoire, A.
  Rare events analysis of temperature chaos in the
  Sherrington-Kirkpatrick model.
  {\it J. Stat. Mech.,} P040016 (2014).

\bibitem{tchea1}
  Kondor, I.
  On chaos in spin glasses.
  {\it J. Phys. A} {\bf 22,} L163--L168 (1989).

\bibitem{tchea2}
  Katzgraber, H.~G. \& Krz\k{a}ka{\l}a, F.
  Temperature and Disorder Chaos in Three-Dimensional Ising Spin
  Glasses.
  {\it Phys. Rev. Lett.} {\bf 98,} 017201 (2007).

\bibitem{barriers}
  Vertechi, D. \& Virasoro, M.A.,
  Enegy barriers in SK spin-glass model.
  {\it J. Phys. France} {\bf 50,} 2325--2332 (1989).

\bibitem{h0}
  Sommers, H.~J. \& Dupont, W.
  Distribution of frozen fields in the mean-field theory of spin glasses.
  {\it J. Phys.} {\bf C17,} 5785--5793 (1984).

\bibitem{venturelli}
  Venturelli, D., Mandr\`a, S., Knysh, S., O'Gorman, B., Biswas, R. \& Smelyanskiy, V.
  Quantum annealing of fully-connected spin glass.
  {\it Phys. Rev. X} {\bf 5,} 031040 (2015).

\bibitem{chimeraT0}
  Katzgraber, H.~G., Hamze, F. \& Andrist R.~S.
  Glassy chimeras may be blind to quantum speedup.
  {\it Phys. Rev. X} {\bf 4,} 021008 (2014).

\bibitem{quantum2d}
  Rieger, H. \& Young, A.~P.
  Zero-temperature quantum phase transition of a two-dimensional Ising
  spin glass.
  {\it Phys. Rev. Lett.} {\bf 72,} 4141--4144 (1994).

\bibitem{farhisa}
  Farhi, E., Goldstone, J. \& Gutmann, S.
  Quantum adiabatic annealing algorithms versus simulated annealing.
  Preprint at http://arxiv.org/abs/quant-ph/0201031 (2002).

\bibitem{algo}
  M\'ezard, M., Parisi, G., Zecchina, R.
  Analytic and algorithmic solution of random satisfiability problems.
  {\it Science} {\bf 297,} 812--815 (2002).

\end{thebibliography}
\end{document}